\documentclass[12pt]{article}
\usepackage{amsmath,bm}
\usepackage{amssymb}
\usepackage{geometry}
\usepackage{graphicx}
\usepackage{hyperref}
\usepackage{subfig}
\usepackage{mciteplus}
\usepackage{float}

\def\gtwid{\mathrel{\raise.3ex\hbox{$>$\kern-.75em\lower1ex\hbox{$\sim
$}}}}
\def\vio{\mathrel{\hbox{$E$\kern-.60em\hbox{$/
$}}}}
\newcommand{\newc}{\newcommand*}

\long\def\begincomment#1\endcomment{%
        \begingroup\sf\baselineskip12pt#1\endgroup}

\newc{\etal}{\textrm{et al.}} 
\newc{\eg}{\textrm{e.g.}} 
\newc{\ie}{\textrm{i.e.}}
\newc{\etc}{\textrm{etc}}
\newc\vs{\textrm{vs.}}
\newc{\cl}{\rm {C.L.}}
\newc{\ev}{\ensuremath{\,\mathrm{eV}}}
\newc{\kev}{\ensuremath{\,\mathrm{keV}}}
\newc{\mev}{\ensuremath{\,\mathrm{MeV}}}
\newc{\gev}{\ensuremath{\,\mathrm{GeV}}}
\newc{\tev}{\ensuremath{\,\mathrm{TeV}}}
\newc{\MeV}{\mev} 
\newc{\TeV}{\tev}
\newc{\invpb}{\ensuremath{/\text{pb}}}
\newc{\invfb}{\ensuremath{\,\textrm{fb}^{-1}}}
\newc\nb{\ensuremath{\,\mathrm{nb}}} \newc\pb{\ensuremath{\,\mathrm{pb}}} \newc\fb{\ensuremath{\,\mathrm{fb}}}
\newc\pc{\ensuremath{\,\mathrm{pc}}}
\newc\kpc{\ensuremath{\,\mathrm{kpc}}}
\newc\mpc{\ensuremath{\,\mathrm{Mpc}}}
\newc\ps{\ensuremath{\,\mathrm{ps}}} 
\newc\cmeter{\ensuremath{\,\mathrm{cm}}} 
\newc\meter{\ensuremath{\,\mathrm{m}}} 
\newc\kmeter{\ensuremath{\,\mathrm{km}}}
\newc\second{\ensuremath{\,\mathrm{s}}}
\newc\msecond{\ensuremath{\,\mathrm{ms}}}
\newc\nsecond{\ensuremath{\,\mathrm{ns}}}
\newc\psecond{\ensuremath{\,\mathrm{ps}}}
\newc{\chisqmin}{\ensuremath{\chi^2_{\mathrm{min}}}}
\newc{\Delchisq}{\ensuremath{\Delta\chi^2}}
\newc{\chisq}{\ensuremath{\chi^2}}
\newc{\like}{\ensuremath{\mathcal{L}}}
\newc\lsim{\ensuremath{\mathrel{\rlap{\lower4pt\hbox{\hskip1pt$\sim$}}\raise1pt\hbox{$<$}}}}
\newc\gsim{\ensuremath{\mathrel{\rlap{\lower4pt\hbox{\hskip1pt$\sim$}}\raise1pt\hbox{$>$}}}}
\newc{\VEV}[1]{\ensuremath{\langle #1 \rangle}}
\newc{\dl}{\ensuremath{\stackrel{\leftarrow}{D}}}
\newc{\dr}{\ensuremath{\stackrel{\rightarrow}{D}}}
\newc{\bcenter}{\begin{center}}    \newc{\ecenter}{\end{center}}
\newc{\bfl}{\begin{flushleft}}    \newc{\efl}{\end{flushleft}}
\newc{\bfr}{\begin{flushright}}    \newc{\efr}{\end{flushright}}

\newc{\bi}{\begin{itemize}}
\newc{\ei}{\end{itemize}}
\newc{\bed}{\begin{description}}
\newc{\eed}{\end{description}}
\newc{\ben}{\begin{enumerate}}
\newc{\een}{\end{enumerate}}

\newc{\be}{\begin{equation}}
\newc{\ee}{\end{equation}}
\newc{\bea}{\begin{eqnarray}}
\newc{\eea}{\end{eqnarray}}
\newc{\bfle}{\begin{flalign}}
\newc{\efle}{\end{flalign}}
\newc{\ra}{\rightarrow}
\newc{\alphas}{\ensuremath{\alpha_s}}
\newc{\alphatwo}{\ensuremath{\alpha_2}}
\newc{\alphaone}{\ensuremath{\alpha_1}}
\newc{\alphai}[1]{\ensuremath{\alpha_{#1}}}
\newc{\alphaem}{\ensuremath{\alpha_{\mathrm{em}}}}
\newc{\alphaeff}{\ensuremath{\alpha_{\mathrm{eff}}}}
\newc{\sineff}{\ensuremath{\sin \theta_{\mathrm{eff}}}}
\newc{\sinsqeff}{\ensuremath{\sin^2 \theta_{\mathrm{eff}}}}
\newc{\dalphahad}{\ensuremath{\Delta \alpha_{\mathrm{had}}}}
\newc{\yt}{\ensuremath{h_t}} \newc{\yb}{\ensuremath{h_b}} \newc{\ytau}{\ensuremath{h_{\tau}}}
\newc\mz{\ensuremath{M_Z}} 
\newc\mw{\ensuremath{m_W}}
\newc\mZ{\mz}        \newc\mW{\mw}
\newc\mhsm{\ensuremath{ m_{H_{\mathrm{SM}}}}}
\newc{\mtop}{\ensuremath{ m_t}}               \newc{\mtpole}{\ensuremath{ M_t}}
\newc{\mbottom}{\ensuremath{ m_b}} 
\newc{\mtau}{\ensuremath{ m_{\tau}}}
\newc{\mt}{\mtpole}
\newc{\mb}{\mbottom} 
\newc{\rtwogg}{\ensuremath{R_{h_2}(\gamma\gamma)}}
\newc{\rtwozz}{\ensuremath{R_{h_2}(ZZ)}}
\newc{\ronegg}{\ensuremath{R_{h_1}(\gamma\gamma)}}
\newc{\ronezz}{\ensuremath{R_{h_1}(ZZ)}}
\newc{\rsiggg}{\ensuremath{R_{h_\textrm{sig}}(\gamma\gamma)}}
\newc{\rsigzz}{\ensuremath{R_{h_\textrm{sig}}(ZZ)}}
\newc{\llbar}{\ensuremath{\ell\bar{\ell}}}
\newc{\tauptaum}{\ensuremath{ \tau^+\tau^-}}
\newc{\qqbar}{\ensuremath{ q\bar{q}}} \newc{\ppbar}{\ensuremath{ p\bar{p}}}
\newc{\bbbar}{\ensuremath{ b\bar{b}}} \newc{\ttbar}{\ensuremath{ t\bar{t}}}
\newc{\ffbar}{\ensuremath{ f\bar{f}}} \newc{\tautaubar}{\ensuremath{ \tau\bar{\tau}}}

\newc{\mchi}{\ensuremath{m_\neutone}}
\newc{\squark}{\ensuremath{\tilde{q}}}
\newc{\slepton}{\ensuremath{\tilde{l}}}
\newc{\gluino}{\ensuremath{\tilde{g}}} 
\newc{\mgluino}{\ensuremath{{m_{\gluino}}}}
\newc{\wino}{\ensuremath{\tilde{W}}} 
\newc{\mwino}{\ensuremath{{m_{\wino}}}}
\newc{\tone}{\ensuremath{{\tilde{t}_1}}}
\newc{\bone}{\ensuremath{{\tilde{b}_1}}}
\newc{\Hone}{\ensuremath{{\tilde{H}_{1}}}}
\newc{\Htwo}{\ensuremath{{\tilde{H}_{2}}}}
\newc{\Hhtwo}{\ensuremath{{H_{2}}}}
\newc{\qli}{\ensuremath{{\tilde{Q}_{i}}}}
\newc{\uri}{\ensuremath{{\tilde{u}_{i}}}}
\newc{\dri}{\ensuremath{{\tilde{d}_{i}}}}
\newc{\lli}{\ensuremath{{\tilde{L}_{i}}}}
\newc{\eri}{\ensuremath{{\tilde{e}_{i}}}}
\newc{\sthw}{\ensuremath{ \sin\theta_W}}              \newc{\cthw}{\ensuremath{\cos\theta_W}}
\newc{\tanthw}{\ensuremath{ \tan\theta_W}}              \newc{\cotthw}{\ensuremath{\cot\theta_W}}
\newc{\ssqthw}{\ensuremath{\sin^2 \theta_W}}
\newc{\msbar}{\ensuremath{\overline{MS}}} \newc{\drbar}{\ensuremath{\overline{DR}}}
\newc{\mtmtsmmsbar}{\ensuremath{ m_t(m_t)^{\msbar}_{{\mathrm{SM}}}}}
\newc{\mtmtsmdrbar}{\ensuremath{ m_t(m_t)^{\drbar}_{{\mathrm{SM}}}}}
\newc{\mtmtmssmdrbar}{\ensuremath{ m_t(m_t)^{\drbar}_{{\mathrm{SUSY}}}}}
\newc{\mbmbmsbar}{\ensuremath{ m_b(m_b)^{\msbar} }}
\newc{\mbmbsmmsbar}{\ensuremath{ m_b(m_b)^{\msbar}_{{\mathrm{SM}}}}}
\newc{\mbmzsmmsbar}{\ensuremath{ m_b(\mz)^{\msbar}_{{\mathrm{SM}}}}}
\newc{\mbmzsmdrbar}{\ensuremath{ m_b(\mz)^{\drbar}_{{\mathrm{SM}}}}}
\newc{\mbmzmssmdrbar}{\ensuremath{ m_b(\mz)^{\drbar}_{{\mathrm{SUSY}}}}}
\newc{\mtaumzsmmsbar}{\ensuremath{ m_{\tau}(\mz)^{\msbar}_{{\mathrm{SM}}}}}
\newc{\mtaumzsmdrbar}{\ensuremath{ m_{\tau}(\mz)^{\drbar}_{{\mathrm{SM}}}}}
\newc{\mtaumzmssmdrbar}{\ensuremath{ m_{\tau}(\mz)^{\drbar}_{{\mathrm{SUSY}}}}}
\newc{\alphasmzms}{\ensuremath{\alpha_s(M_Z)^{\overline{MS}}}}
\newc{\alphaimzms}[1]{\ensuremath{\alpha_{#1}(M_Z)^{\overline{MS}}}}
\newc{\alphaemmz}{\ensuremath{\alpha_{\mathrm{em}}(M_Z)^{\overline{MS}}}}
\newc{\mzero}{\ensuremath{{m_0}}}
\newc{\mhalf}{\ensuremath{ m_{1/2}}}
\newc{\tanb}{\ensuremath{\tan\beta}}
\newc{\azero}{\ensuremath{ A_0}}
\newc{\signmu}{\ensuremath{\rm{sgn}\,\mu}}
\newc{\atau}{\ensuremath{{A_{\tau}}}}
\newc{\mueff}{\ensuremath{\mu_{\rm{eff}}}}
\newc{\lam}{\ensuremath{{\lambda}}}
\newc{\kap}{\ensuremath{{\kappa}}}
\newc{\alam}{\ensuremath{{A_{\lambda}}}}
\newc{\akap}{\ensuremath{{A_{\kappa}}}}
\newc{\hs}{\ensuremath{ H_s}}      
\newc{\mhs}{\ensuremath{ m_{H_s}}} 
\newc{\mgut}{\ensuremath{ M_{\rm GUT}}}
\newc{\gut}{\ensuremath{{\rm GUT}}}
\newc{\mplanck}{\ensuremath{ M_{\rm P}}}      \newc{\mpl}{\ensuremath{ M_{\rm Pl}}}
\newc{\msusy}{\ensuremath{ M_{\rm SUSY}}}      \newc{\ms}{\ensuremath{ M_{\rm S}}}
 \newc{\hu}{\ensuremath{ H_u}}       \newc{\hd}{\ensuremath{ H_d}}
 \newc{\mhu}{\ensuremath{ m_{H_u}}}       \newc{\mhd}{\ensuremath{ m_{H_d}}}
 \newc{\mhuew}{\ensuremath{ m^{\ast}_{H_u}}}       \newc{\mhdew}{\ensuremath{ m^{\ast}_{H_d}}}
 \newc{\mhuewsq}{\ensuremath{ m^{\ast\, 2}_{H_u}}}       \newc{\mhdewsq}{\ensuremath{ m^{\ast\, 2}_{H_d}}}
 \newc{\mhl}{\ensuremath{m_\hl}} 
 \newc{\mhone}{\ensuremath{m_{h_1}}} 
 \newc{\mhtwo}{\ensuremath{m_{h_2}}} 
 \newc{\mhi}{\ensuremath{m_{\tilde{h}}}} 
 \newc{\mul}{\ensuremath{m_{\tilde{u}_L}}} 
 \newc{\mbone}{\ensuremath{m_{\tilde{b}_1}}}  
 \newc{\mtone}{\ensuremath{m_{\tilde{t}_1}}} 
 \newc{\ma}{\ensuremath{m_A}} 
 \newc{\mH}{\ensuremath{m_H}} 
 \newc{\maone}{\ensuremath{m_{a_1}}} 
 \newc{\matwo}{\ensuremath{m_{a_2}}}
 \newc{\hone}{\ensuremath{h_1}}
 \newc{\htwo}{\ensuremath{h_2}}
 \newc{\aone}{\ensuremath{a_1}}
 \newc{\atwo}{\ensuremath{a_2}}
 \newc{\mqthree}{\ensuremath{m_{\tilde{Q}_3}^2}}
 \newc{\muthree}{\ensuremath{m_{\tilde{u}_3}^2}}
 \newc{\mqli}{\ensuremath{m_{\tilde{Q}_{i}}}}
 \newc{\muri}{\ensuremath{m_{\tilde{u}_{i}}}}
 \newc{\mdri}{\ensuremath{m_{\tilde{d}_{i}}}}
 \newc{\mlli}{\ensuremath{m_{\tilde{L}_{i}}}}
 \newc{\meri}{\ensuremath{m_{\tilde{e}_{i}}}}
 \newc{\ts}{\ensuremath{T_{SUSY}}}

\newc{\sigsip}{\ensuremath{\sigma^{\rm SI}_{p}}}	\newc{\sigsin}{\ensuremath{\sigma^{\rm SI}_{n}}}
\newc{\sigsdp}{\ensuremath{\sigma^{\rm SD}_{p}}}	\newc{\sigsdn}{\ensuremath{\sigma^{\rm SD}_{n}}}
\newc{\sigsi}{\ensuremath{\sigma^{\rm SI}}}	\newc{\sigsd}{\ensuremath{\sigma^{\rm SD}}}
\newc{\abund}{\ensuremath{ \Omega h^2}}
\newc{\omegadm}{\ensuremath{ \Omega_{{\rm DM}}}}     \newc{\abunddm}{\ensuremath{ \Omega_{{\rm DM}} h^2}} 
\newc{\omegam}{\ensuremath{ \Omega_{{\rm m}}}}       \newc{\abundm}{\ensuremath{ \Omega_{{\rm m}} h^2}}
\newc{\omegab}{\ensuremath{ \Omega_{{\rm b}}}}	\newc{\abundb}{\ensuremath{ \Omega_{{\rm b}} h^2}}
\newc{\omegatot}{\ensuremath{ \Omega_{{\rm TOT}}}}
\newc{\omegacdm}{\ensuremath{ \Omega_{{\rm CDM}}}}   \newc{\abundcdm}{\ensuremath{ \Omega_{{\rm CDM}} h^2}}
\newc{\omegalambda}{\ensuremath{ \Omega_{\Lambda}}} \newc{\abundlambda}{\ensuremath{ \Omega_{\Lambda} h^2}}
\newc{\omegarad}{\ensuremath{ \Omega_{{\rm rad}}}}  \newc{\abundrad}{\ensuremath{ \Omega_{{\rm rad}} h^2}}
\newc{\rhocrit}{\ensuremath{ \rho_{\rm crit}}}
\newc{\rhochi}{\ensuremath{ \rho_{\chi}}}
\newc{\abunchi}{\ensuremath{\Omega_\chi h^2}}
\newc{\abundlsp}{\ensuremath{\Omega_{\rm LSP}h^2}}
\newc{\amu}{\ensuremath{ a_{\mu}}}        \newc{\amususy}{\ensuremath{ a_{\mu}^{\mathrm{SUSY}}}}
\newc{\amuexpt}{\ensuremath{ a_{\mu}^{\mathrm{expt}}}}        \newc{\amusm}{\ensuremath{ a_{\mu}^{\mathrm{SM}}}}
\newc\deltaamu{\ensuremath{\Delta a_{\mu}}} \newc{\deltaamususy}{\ensuremath{\delta a_{\mu}^{\mathrm{SUSY}}}}
\newc\gmtwo{\ensuremath{ (g-2)_{\mu}}} 
\newc{\deltagmtwomususy}{\ensuremath{\delta\left(g-2\right)_{\mu}^{\mathrm{SUSY}}}}
\newc{\deltagmtwomu}{\ensuremath{\delta\left(g-2\right)_{\mu}}}
\newc\BR{\ensuremath{\rm BR}}
\newc\bsgamma{\ensuremath{ b\rightarrow s \gamma }}
\newc\bxsgamma{\ensuremath{\overline{B}\rightarrow X_{s}\gamma}}
\newc\brbsgamma{\ensuremath{\BR\left(\bsgamma\right)}}
\newc\brbxsgamma{\ensuremath{\BR\left(\bxsgamma\right)}}
\newc\bsmumu{\ensuremath{B_s\to\mu^+\mu^-}}
\newc\brbsmumu{\ensuremath{\BR\left(B_s\to\mu^+\mu^-\right)}}
\newc\bdmmumu{\ensuremath{\overline{B}_d\to\mu^+\mu^-}}
\newc\bbbarmix{\ensuremath{\overline{B}_s\mbox{-}B_s}}      % B_s mixing
\newc\delmbs{\ensuremath{\Delta M_{B_s}}}
\newc{\butaunu}{\ensuremath{B_u \rightarrow \tau \nu}}
\newc{\brbutaunu}{\ensuremath{\BR\left(B_u \rightarrow \tau \nu\right)}}
\newcommand*{\reftable}[1]{Table~\ref{#1}}         
       
\newcommand*{\reffig}[1]{Fig.~\ref{#1}}
 
        \newcommand*{\refeq}[1]{Eq.~(\ref{#1})}
     \newcommand*{\refsec}[1]{Sec.~\ref{#1}}

\newcommand*{\neutone}{\ensuremath{\chi^0_1}}

\newcommand*{\charone}{\ensuremath{{\chi}^{\pm}_1}}

\newcommand*{\softsusy}{SOFTSUSY}
\newcommand*{\feynhiggs}{FeynHiggs}
\newcommand*{\micromegas}{MicrOMEGAs}

\newcommand*{\susyhit}{\text{SUSY-HIT}}

\newcommand*{\superiso}{\text{SuperIso}}

\let\oldcite\cite
\renewcommand*{\cite}{~\oldcite}

\newcommand*{\hl}{\ensuremath{h}}

\newcommand*{\met}{E_T^{\textrm{miss}}}

\newcommand*{\dfi}{\Delta\phi(\met,\textrm{jet})}
\begin{document}

\begin{flushright}
DO-TH 16/30 \\
QFET-2016-17
\end{flushright}

\title{MSSM fits to the ATLAS 1 lepton excess”}  
\author{Kamila Kowalska$^1$ and Enrico Maria Sessolo$^2$\\[2ex]
\small\it $^1$ Fakult\"at f\"ur Physik, TU Dortmund, Otto-Hahn-Str.4, D-44221 Dortmund, Germany\\
\small\it $^2$ National Centre for Nuclear Research, Ho{\. z}a 69, 00-681 Warsaw, Poland \\
}
\date{}

{\let\newpage\relax\maketitle}
%\maketitle
\centering
\url{kamila.kowalska@tu-dortmund.de}\\
\url{enrico.sessolo@ncbj.gov.pl}

\abstract{ We use the framework of the p19MSSM to perform a fit to the mild excesses over the Standard Model background recently observed in three bins of the ATLAS~1-lepton + (b-)jets + $\met$ search. We find a few types of spectra that can fit the emerging signal and at the same time are not excluded by other LHC searches. They can be grouped roughly in two categories. The first class is characterized by the presence of one stop or stop and sbottoms with mass in the ballpark of $700-800\gev$ and a neutralino LSP of mass around $400\gev$, with or without the additional presence of an intermediate chargino. In the second type of scenarios the stop, lightest chargino, sbottom if present, and the neutralino are about or heavier than $\sim 650\gev$ and the signal originates from cascade decays of squarks of the 1st and 2nd generation, which should have a mass of $1.1-1.2\tev$. For the best-fit scenarios, we compare the global chi-squared with respect to several ATLAS and CMS searches with the corresponding chi-squared of the Standard Model expectation, showing that the putative signal is also favored globally with respect to the background only hypothesis. 
We point out that if the observed excess persists in the next round of data, it should be accompanied by associated significant excesses in all-hadronic final state searches.
}
%%%%%%%%%%%%%%%%%%%%%%%%%%%%%%%%%%%%%%%%%%%%%%%%%%%%%%%%%%
\newpage
\section{Introduction}\label{intro}
The LHC is now in the course of its second run, characterized by proton-proton collisions at the center-of-mass energy of 13\tev. 
By August 2016, the ATLAS and CMS Collaborations had collected approximately 14\invfb\ of data and presented the results of their
searches for new physics beyond the Standard Model (BSM) at the ICHEP conference\cite{ichep16} in Chicago.
To some disappointment, in almost all of the hundreds of presented channels
the number of observed events turned out to be in excellent agreement with the Standard Model (SM) expectation, 
so that no new physics seems to be present at the tested energy. 

On the other hand, as is bound to happen given the enormous number of kinematical bins and 
final-state channels analyzed by the two collaborations to cover the parameter space of as many BSM models as possible, 
a few anomalies have emerged to small significance. 
In particular, the 1-lepton ATLAS search\cite{ATLAS-CONF-2016-050} shows a $3.3\,\sigma$ 
excess of events in the so-called DM-low bin, and less significant $2.6\,\sigma$ and $2.2\,\sigma$
excesses in the bins called bC2x-diag and SR1, respectively. 
In this letter we dedicate some attention to these excesses, as they are appearing in one of the classic topologies 
designed to discover supersymmetry (SUSY), which arguably remains the most comprehensive and appealing scenario for BSM physics. 

It is obviously extremely premature to get excited about an excess like this, as its statistical significance 
is relatively low and experience has shown time and again that background fluctuations of comparable strength 
do happen with some frequency, especially when many channels are analyzed. 
On the other hand, given the appeal that SUSY has held on the particle community for many years,
we also think it is legitimate to wonder whether the ATLAS 1-lepton
excess already points toward specific SUSY spectra, and in particular spectra more involved 
than the simplified models used by the experimental collaborations to interpret their results.

We address the question by performing a global fit to the bins showing the excess in the generic 
framework of the phenomenological Minimal Supersymmetric Standard Model (p19MSSM)\cite{Berger:2008cq}, 
which is characterized by 19 independent parameters. 
We use a set of model points generated by the ATLAS Collaboration in Ref.\cite{Aad:2015baa}, 
which satisfies a set of experimental constraints from the relic density, direct dark matter (DM) searches, Higgs sector measurements, 
electroweak and flavor physics, as well as includes the exclusion bounds from the 8\tev\ LHC SUSY searches.
We also use the state-of-the-art exclusion bounds from 11 other ATLAS SUSY searches based on 3.2\invfb\ and 
$\sim$14\invfb\ data, which were derived by one of us in Ref.\cite{Kowalska:2016ent}.
We repeat that, while our analysis satisfies a genuine curiosity,
the reader will have to keep in mind that it might prove a futile effort were 
the excess to disappear or lose significance in the next round of data.   

It is important to note in this regard that the CMS search\cite{CMS-PAS-SUS-16-028} equivalent to 
ATLAS 1-lepton does not show any significant excess over the SM background with $\sim$13\invfb\ of data. 
 By recasting the CMS analysis we will show, however, that this is not in 
contradiction with the onset of a possible signal observed at ATLAS, as the kinematical variables used by CMS 
and the subsequent definitions of the signal bins differ from the ones of ATLAS 
and the compatibility between the two is not straightforward.

Recently, an analysis of the ATLAS 1-lepton excess was performed in Ref.\cite{Han:2016hgr}. 
The excess was fitted with simplified SUSY model spectra (SMS) characterized by moderately light stops decaying
into a bino or, in alternative, higgsino lightest SUSY particle (LSP).  The originating signal was 
confronted with two CMS hadronic searches\cite{CMS-PAS-SUS-16-029,CMS-PAS-SUS-16-030}. 
It was concluded in\cite{Han:2016hgr} that the signal is consistent at $2\sigma$ with
a spectrum characterized by a $\sim 750-800\gev$ stop, a bino LSP and a higgsino next-to-LSP.
We extend the analysis here by considering a broader range of possibilities for the SUSY spectrum, as embodied by our choice 
of using the p19MSSM points to fit the excess. We show that there exist additional 
spectra and decay chains that can equally well fit the 1-lepton excess, 
and for each found scenario we compare its global $\chi^2$ with respect to a sample of LHC searches with the $\chi^2$ of
the background only hypothesis, showing that all our scenarios are favored globally over the SM to some significance.  
We additionally provide possible signals for the next batch of LHC data and
give a few comments on possible UV completions.

 This paper is organized as follows. In \refsec{num} we outline the fitting procedure and we enumerate the constraints we 
impose on our model spectra. In \refsec{res} we present and discuss the main results of the fit and favored SUSY spectra. 
In \refsec{global} we evaluate the significance of the best-fit spectra globally with respect to the background only hypothesis.
We summarize our findings and conclude in \refsec{sum}. Additionally, we present two appendices dedicated, respectively, to the validation of the CMS 1-lepton search, and to the detailed breakdown of the chi-squared contributions of every bin in the most important searches considered in this paper.

%%%%%%%%%%%%%%%%%%%%%%%%%%%%%%%%%%%%%%%%%%%%%%%%%%%%%%%%%%
\section{Fitting procedure and constraints}\label{num}

We investigate in this paper whether the excess in events shown by the ATLAS 1~lepton + (b-)jets 
search\cite{ATLAS-CONF-2016-050} can be interpreted in 
the generic framework of the p19MSSM.

To fit the three excesses we use a set of model points provided by the ATLAS Collaboration in Ref.\cite{Aad:2015baa} 
and generated using methods similar to those presented in\cite{Berger:2008cq,CahillRowley:2012cb,CahillRowley:2012kx,Cahill-Rowley:2014twa}. 
The set consists of points for which mass and trilinear parameters have been scanned up to a maximum value of 4\tev, with the exception of the third generation trilinear soft coupling $A_t$, scanned in the range $[-8\tev,8\tev]$ to allow for the correct Higgs mass even when the stop mass is not very large. 
Phenomenological constraints from the relic density (upper bound), DM searches, Higgs and flavor physics, 
and electroweak precision data were taken into account,  
as was the impact of 22 ATLAS LHC searches at $\sqrt{s}=7$ and $8\tev$ with integrated luminosity of  20.3\invfb. 
The SUSY spectra were calculated with \softsusy\ v.3.4.0\cite{Allanach:2001kg}, and the decay branching ratios with \susyhit\ v.1.3\cite{Djouadi:2006bz}.
\micromegas\ v.3.5.5\cite{Belanger:2006is,Belanger:2010pz} and \superiso\cite{Mahmoudi:2008tp} were used to evaluate dark matter, precision electroweak and flavor observables, while the lightest Higgs boson mass was calculated with \feynhiggs\ v.2.10.0\cite{Heinemeyer:1998yj,Hahn:2013ria}.
The resulting p19MSSM set contained 183,030 allowed model points. More details on the sampling procedure and implementation of the experimental bounds
can be found in Ref.\cite{Aad:2015baa}.

We proceed as follows. We start with the sample of points that is not at present excluded by the 
current ATLAS~13\tev\ searches for SUSY particles, which was constructed in Ref.\cite{Kowalska:2016ent} out of the 
183,030 initial points.\footnote{Other papers analyzing the impact of the most recent LHC bounds  
on natural SUSY spectra can be found in\cite{Han:2016xet,Buckley:2016kvr}.} 
All points were there tested on-the-fly by recasting for the p19MSSM 11 direct ATLAS searches:
\begin{itemize}
\item ATLAS 0 leptons + 2-6 jets + $\met$, 3.2\invfb\cite{Aaboud:2016zdn}, 13.3\invfb\cite{ATLAS-CONF-2016-078},
\item ATLAS 1 lepton + jets + $\met$, 3.2\invfb\cite{Aad:2016qqk}, 14.8\invfb\cite{ATLAS-CONF-2016-054},
\item ATLAS 3 b-tagged jets + $\met$, 3.2\invfb\cite{Aad:2016eki}, 14.8\invfb\cite{ATLAS-CONF-2016-052},
\item ATLAS 0 leptons + (b-)jets + $\met$, 13.3\invfb\cite{ATLAS-CONF-2016-077},
\item ATLAS 1 lepton + (b-)jets + $\met$, 3.2\invfb\cite{Aaboud:2016lwz},
\item ATLAS 2 leptons + jets + $\met$, 3.2\invfb\cite{ATLAS-CONF-2016-009},
\item ATLAS 2 b-tagged jets + $\met$, 3.2\invfb\cite{Aaboud:2016nwl},
\item ATLAS monojet + $\met$, 3.2\invfb\cite{Aaboud:2016tnv}.
\end{itemize}
Detailed description of the recast procedure, as well as validations of the implemented searches, 
can be found in Ref.\cite{Kowalska:2016ent}.

Additionally, there are a few CMS~13\tev\ SUSY searches that should be taken into account 
as they provide results complementary to those of ATLAS, in particular the 1-lepton\cite{CMS-PAS-SUS-16-028} and
0-lepton\cite{CMS-PAS-SUS-16-014,CMS-PAS-SUS-16-015,CMS-PAS-SUS-16-016} searches. 
Reference\cite{CMS-PAS-SUS-16-028} 
is similar to the ATLAS search presented in\cite{ATLAS-CONF-2016-050} but, in contrast to the ATLAS results, it 
does not show any significant excess in the kinematical bins. 
As was mentioned in \refsec{intro}, this does not automatically reflect an inconsistency between the two results,
as the definitions of the bins in both searches are quite different, but one needs to make sure that the models that 
fit the ATLAS excess are not instead excluded by CMS.
To evaluate the impact of the CMS 1-lepton search\cite{CMS-PAS-SUS-16-028} on our model sample, we implemented it in our recast tool as well.
The validation of the recast procedure against the experimental results is given in Appendix \ref{CMS_valid}.

% One-lepton searches typically involve the tagging of a $W$ boson originating from the decay of a top quark and, as a consequence, 
% they are sensitive to the $\tilde{t}_1\to t \neutone$ and $\tilde{b}_1\to t \charone$ decay chains. 
% Therefore, if the branching ratio into a top quark final state is reduced the exclusion bounds will become weaker.
% To evaluate the impact of the CMS search\cite{CMS-PAS-SUS-16-028} on our model sample, 
% we compare the selected points with the exclusion bound presented by CMS in the framework of their SMS of choice 
% (stop pair production with $\textrm{BR}(\tilde{t}_1\to t \neutone)=100\%$).
% For the points that happen to lie inside the CMS 95\%~C.L. exclusion contour 
% we consider the branching ratio into the $t \neutone$ 
% final state. 
% In the case of the ATLAS 1-lepton search 
% we have heuristically determined the patterns relating decay branching ratios and exclusion bounds 
% based on the numerical simulations presented in Ref.\cite{Kowalska:2016ent}. 
% For example, because of the veto on the absence of a hard lepton in the final state, assuming 
% $\textrm{BR}(\tilde{t}_1\to t \neutone)=50\%$ (with the remaining 50\% corresponding to all-hadronic final states)
% leads to a weakening of the bound on the stop mass by about 50\gev\ for a wide range of neutralino mass choices.
% We assume in this work that relations between branching ratio and exclusion bound similar to ATLAS hold 
% for the CMS search as well.

The 95\%~C.L. bounds on the stop/neutralino, sbottom/neutralino SMS provided in the CMS 0-lepton analyses 
are in general more constraining than those of the equivalent ATLAS searches\cite{ATLAS-CONF-2016-078,ATLAS-CONF-2016-077} 
in the part of the parameter space where the ATLAS 1-lepton excess can be fitted.
Bounds obtained in 0-lepton searches tend to be fairly stable under variations of the  
decay channels, so that for most of our points we will take the CMS exclusion lines at face value. 
On the other hand, we will draw attention to a few spectra that could fit the ATLAS excess if they were
not excluded by the CMS 0-lepton searches when the SMS bound 
is taken at face value. Those spectra are characterized by a significant reduction of the branching ratio to all-hadronic final states, 
so that there is some doubt on whether they are actually excluded or not, and the answer would require a detailed recasting of 
the CMS searches as was done in Ref.\cite{Kowalska:2016ent} for ATLAS. 
We refrain from doing this here, given the still premature nature of the ATLAS excess, 
and limit ourselves to pointing out in the text the presence of these points ``of doubt.''

%%%%%%%%%%%%%%%%%%%%%%%%%%%%%%%%%%%%%%%%%%%%%%
\begin{table}[t]
\begin{centering}
\begin{tabular}{|c|c|c|c|c|}
\hline 
 Bin & Obs. & Bkg. & Sig. & $\sigma$ ($p$-value)\\
\hline 
 SR1 & 37 & $24\pm 3$ & 13.2 & 2.2 (0.012)\\
 bC2x-diag & 37 & $22\pm 3$ & 15.2 & 2.6 (0.004) \\
 DM-low & 35 & $17\pm 2$ & 18.1 & 3.3 (0.0004) \\
\hline
\end{tabular}
\caption{Summary of the excesses observed in the ATLAS 1 lepton + (b-)jets + $\met$ search\cite{ATLAS-CONF-2016-050}. 
The name of the bin, the number of observed and background events, as well as the significance of the excess are provided by the experimental collaboration. The number of signal events that fit the excess best is
derived according to the procedure described in the text.}
\label{tab:ATLAS_LHC_13TeV_excesses}
\end{centering}
\end{table}
%%%%%%%%%%%%%%%%%%%%%%%%%%%%%%%%%%%%%%%%%%%%%%%%%%%

We then proceed to perform a maximum likelihood estimate for the ATLAS 1~lepton + (b-)jets + $\met$ search.
A moderate excess shows up in 3 of the experimental bins\cite{ATLAS-CONF-2016-050}, see \reftable{tab:ATLAS_LHC_13TeV_excesses}.
The bins are not statistically independent, 
and the same signal could be responsible for all 3 observed excesses. We limit ourselves to this hypothesis in this paper. 

Indicating the observed number of events, the expected SM background, and the collaboration's estimate of the
systematic uncertainty for each bin $i$ as $o_i$, $b_i$, and $\delta b_i$, respectively, we seek the maximum 
of the likelihood function      
\begin{equation}\label{eq:like}
\mathcal{L}_i=\frac{1}{\sqrt{2\pi}\,\delta b_i}\int db'\,e^{-\frac{(b'-b_i)^2}{2\,\delta b_i^2}}\times\frac{e^{-(s_i+b')}(s_i+b')^{o_i}}{o_i!}\,,
\end{equation}
where $s_i$ is the simulated signal for each bin, as a function of the points in the parameter space.
The signals $s_i$ are simulated in the usual way, following the procedure given, e.g., in\cite{Kowalska:2016ent}.
We use \texttt{PYTHIA~8}\cite{Sjostrand:2007gs} for the hard-scattering event and showering and \texttt{DELPHES~3}\cite{deFavereau:2013fsa} 
for the detector properties' simulation.
We fit to all bins by making use of the $\Delta\chisq$ variable: $\Delta\chisq_i=-2\log(\mathcal{L_{\textrm{i}}}/\mathcal{L}_{0})$, where
$\mathcal{L}_{0}$ corresponds to the background-only hypothesis for the bins that do not show an excess, and to the signal given in \reftable{tab:ATLAS_LHC_13TeV_excesses}
otherwise. Since the kinematical bins considered in the ATLAS 1-lepton search are not independent
we build a global statistics as
\be\label{eq:chitot}
\Delta\chisq_{1\textrm{-lep}}=\max_i\,\Delta\chisq_i\,.
\ee

The sample of viable points fitting the excesses is narrowed down in a two-step procedure. 
First we identify the points that fit the excesses approximately at the 95\%~C.L. In other words we require 
$\Delta\chisq_{1\textrm{-lep}}\leq 3.84$.  
Given that the most significant excess in \reftable{tab:ATLAS_LHC_13TeV_excesses} is at most 
$3.3\,\sigma$, this first step allows a large number of points to survive. 
To further scale down the sample, we assume that the largest 
excess will yield in the near future a $5\,\sigma$ discovery. 
A simple rescaling of the integrated luminosity gives us an approximate estimate of the luminosity $L'$ at which this might occur.
Equating $\sqrt{L'/L_0}=5\,\sigma/3.3\,\sigma$ yields, given the present luminosity of $L_0=13.2\invfb$, 
a target $L'\approx 32\invfb$.  
Thus, as a final step, we rescale $o_i$ and $b_i$ in the 3 bins presenting an excess by the target luminosity $L'$ and subsequently 
fit to the maximum likelihood of \refeq{eq:like} at the 95\%~C.L.
We present in the next section the best fit spectra.

%%%%%%%%%%%%%%%%%%%%%%%%%%%%%%%%%%%%%%%%%%%%%%%%%%%%%%%%%%
\section{Spectra that fit the signal}\label{res}

Out of 183,030 p19MSSM model points allowed by the 8\tev\ LHC data, 138,141 are not excluded after inclusion of the 13\tev\ results. 
1,293 model points fit the excess observed by ATLAS~1~lepton + (b-)jets + $\met$ $13.2\invfb$ search at the 95\%~C.L., 
while 194 would produce a $5\sigma$ signal at the luminosity of $\sim 32\invfb$. 
Out of those, 48  points are not excluded when the CMS 1- and 0-lepton searches are taken into account. 
These will be considered in what follows as the allowed models.

\begin{figure}[t]
\centering
\subfloat[]{
\label{fig:a}
\includegraphics[width=0.75\textwidth]{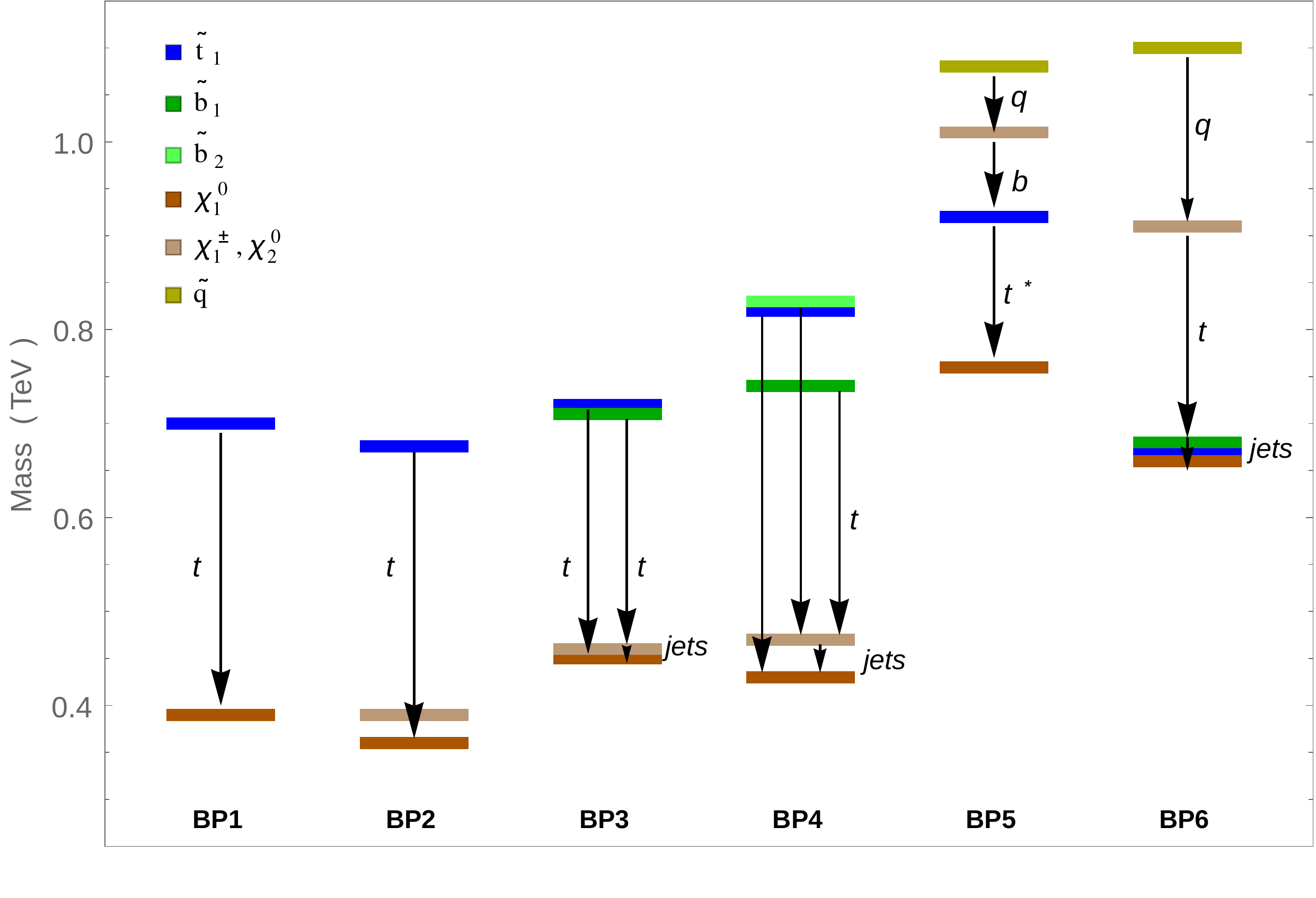}
}
\caption{Benchmark spectra that can fit the ATLAS~1~lepton + (b-)jets + $\met$ excesses\cite{ATLAS-CONF-2016-050}. 
The colors are associated with each sparticle as described in the legend.
We indicate the SM particle emitted at every step of the decay chain next to the corresponding arrow.
}
\label{fig:spectra}
\end{figure}

The resulting spectra can be divided into a few categories, 
depending on the properties of the light particles present in the spectrum and the mechanism that allows to 
fit the observed excesses. We describe them below, and summarize their features in \reftable{tab:benchmarks} and in 
\reffig{fig:spectra}.\bigskip

%%%%%%%%%%%%%%%%%%%%%%%%%%%%%%%%%%%%%%%%%%%%%%%%%%%%%%%
\begin{table}[t]
 \centering
 \begin{tabular}{|c|c|c|c|c|c|c|c|} 
 \hline
 Benchmark & BP1 & BP2 & BP3 & BP4 & BP5 & BP6 \\
 \hline
 $m_{\tilde{t}_1}$  & 700 & 676 & 700 & 829 &  924 & 677  \\
 $m_{\tilde{b}_1}$ & 1100 & 1665 & 700 &  738 & 2259 &683 \\
 $m_{\tilde{b}_2}$ & 2080 & 3245 & 3650 &  834  & 3227 &2050 \\ 
 $m_{\tilde{q}_{L}}$ & 2520 & 2545  & 3630 & 3850  & 1080 & 1100 \\
 $m_{\chi_1^\pm}$, $m_{\chi_2^0}$ & 896 & 390& 460 & 463 & 1010 & 914\\
 $m_{\chi_1^0}$ & 386 & 369 & 453 & 434 & 760 & 665\\
 \hline
 $\textrm{BR}(\tilde{t}_1\to t \neutone)$ & 100\% & 72\% & 87\% & 82\%&  &\\
 $\textrm{BR}(\tilde{b}_1\to t\charone)$ & & & 86\% & 28\% &  & \\
 $\textrm{BR}(\tilde{b}_2\to t\charone)$ & & &  & 82\% &  &  \\
 $\textrm{BR}(\tilde{q}\to q\charone\to q\,b\tilde{t}_1)$ & & &  & & 56\% & \\
 $\textrm{BR}(\tilde{q}\to q\charone\to q\,t\tilde{b}_1)$ & & &  & & & 38\% \\
  \hline
 $\Delta\chisq_{1\textrm{-lep}}$ & 0.6 & 1.0 & 2.0 & 1.3 & 0.9 & 0.5 \\
 \hline
 \end{tabular}
 \caption{Benchmark point models that fit the excess observed by the ATLAS~1~lepton + (b-)jets + $\met$ search\cite{ATLAS-CONF-2016-050}.
The mass of the light degrees of freedom in\gev, the branching ratio of the dominant decay yielding a hard lepton,
and the goodness of fit according to \refeq{eq:chitot} are shown.}
\label{tab:benchmarks}
\end{table}
%%%%%%%%%%%%%%%%%%%%%%%%%%%%%%%%%%%%%%%%%%%%%%%%%%%%%%%%%

$\bullet$ {\it Right-chiral stop scenario.}
The signal is consistent with the SMS employed by the experimental collaboration or similar spectra.
The signal can be generated by the pair-produced stops decaying into a top quark and the LSP with 
$\textrm{BR}(\tilde{t}_1\to t \neutone)\approx 100\%$, as exemplified in \reftable{tab:benchmarks} by the benchmark BP1.
The table shows also the branching ratio of the mother sparticle into the channel that produces a hard lepton, 
and the $\chi^2$ value of this point according to \refeq{eq:chitot} (at the luminosity of $13.2\invfb$).
The stop in this scenario is moderately light and mostly right-chiral, with $m_{\tone}\approx 600-800\gev$,
and the source of the missing energy is a bino-like neutralino LSP with mass in the range $\sim 360-410\gev$. 
Note that spectra similar to this scenario were found to fit the signal in\cite{Han:2016hgr}.

The model points of the ATLAS set\cite{Aad:2015baa} all yield an acceptable value for the relic abundance (at least within
an upper bound), which means that when the LSP is bino-like the spectrum presents also additional light 
particles in mass close to $m_{\chi_1^0}$. In the case of BP1, 
these are a light right-chiral selectron and smuon with just about the same mass as the neutralino, which 
contribute to the early Universe co-annihilation with the LSP. Because their mass is almost degenerate with 
$m_{\chi_1^0}$, direct SUSY searches are not yet sensitive to the presence of these particles, and since they 
do not affect the fitted 1-lepton signal we do not indicate their presence in \reftable{tab:benchmarks} and in 
\reffig{fig:spectra}.

Alternatively, there might be a light chargino (either wino- or higgsino-like) in the spectrum, 
which participates in co-annihilating with the neutralino 
in the early Universe, as exemplified in the spectrum of BP2 in \reftable{tab:benchmarks}. 
Although a light chargino is necessary for the relic density, it does not directly participate to producing a signal in the 
1-lepton ATLAS search. It has the effect of mostly
reducing the branching ratio into the $t \neutone$ final state (because of a non-negligible 
$\textrm{BR}(\tilde{t}_1\to b \charone)$) and thus allowing for a best fit 
involving a slightly lighter stop and neutralino than in BP1.\smallskip

%%%%%%%%%%%%%%%%%%%%%%%%%%%%%%%%%%%%%%%%%%%%%%%%
%%%%%%%%%%%%%%%%%%%%%%%%%%%%%%%%%%%%%%%%%%%%%%%%%%

$\bullet$ {\it Stop/sbottom scenario.} The signal is consistent with the presence of relatively 
light, mostly left-chiral stops and sbottoms, with $m_{\tone}\approx m_{\bone}\approx 650-850\gev$, 
neutralino LSP (bino, higgsino, or an admixture) of mass $\sim 350-480\gev$, and the lightest chargino
or second-lightest neutralino not far above the LSP.
The ATLAS excess is fitted assuming pair-production of left-handed stops and sbottoms, 
with the subsequent decay chain depending on the mass splitting
between $\tilde{t}_1$, $\tilde{b}_1$, and the light electroweakinos. 
Several possibilities are allowed: $\tilde{t}_1\to t\neutone$, $\tilde{t}_1\to t\chi^0_{2,3}$, and 
$\tilde{b}_1\to t\chi_1^{\pm}$, with branching ratios in all cases varying between 50\% and 100\%. 
The charginos and neutralinos just above the LSP decay to $\chi_1^0$ and soft jets via an off-shell gauge boson 
($\textrm{BR}(\chi_2^0,\chi_1^{\pm}\rightarrow \chi_1^0+\textrm{jets})\approx 50-60\%$) so that the acceptance/efficiency
for finding an isolated hard lepton in the event does not drop significantly. 
The benchmark point for this scenario is shown in \reftable{tab:benchmarks} as BP3. 

BP3 is at risk of being excluded by the CMS $\alpha_T$ search\cite{CMS-PAS-SUS-16-016}, as emerges by the interpretation
of this search in the SMS T2bb with a light sbottom and the LSP. 
On the other hand, for BP3 the efficiency of the $\alpha_T$ search to all-hadronic final states is reduced by a factor 2.3 with respect to the T2bb SMS, due to the lepton veto. Only a full simulation of\cite{CMS-PAS-SUS-16-016} can quantitatively assess 
the impact of the efficiency reduction on the 95\%~\cl\ exclusion bound. 
As we explained in \refsec{num}, performing a full simulation of every available search far exceeds the purpose of this paper, 
so that for the moment we limit ourselves to pointing out the existence of this possible tension.

The ATLAS signal can also be fitted when the right-chiral sbottom is light as well. 
In such a case different mass hierarchies can arise between $\tilde{t}_1$ , $\tilde{b}_1$ and $\tilde{b}_2$, 
with all three sparticles contributing to the signal.
The benchmark point for this scenario is shown in \reftable{tab:benchmarks} as BP4.\smallskip

$\bullet$ {\it Squark decay scenario.} 
There exists the possibility of fitting the signal with spectra moderately heavier than those considered so far.
In the case where the right-chiral stop and the LSP lie above $\sim 600\gev$, 
the signal can be fitted in the presence of
squarks of the first two generations with $m_{\tilde{q}_L}\approx 1.1\tev$, which can boost the production cross section.
An example of this light squark spectrum is shown in \reftable{tab:benchmarks} and
\reffig{fig:spectra} as BP5.
The decay chain proceeds through $\tilde{q}_L\to \charone\,q\to b\,\tilde{t}_1\,q$ 
(with approximately 60\% branching ratio)
and the lepton tagged by the ATLAS search originates from the subsequent
decay of the stop.

A variation of the above scenario, involving left-chiral stops/sbottoms instead, is presented as BP6. 
The hard lepton comes now from the decay of the chargino or second-lightest neutralino, as the cascade proceeds through 
$\tilde{q}_L\to \charone (\chi_2^0)\,q\to  t\,\tilde{b}_1(\tilde{t}_1)\,q$, with a final soft decay of 
$\tilde{b}_1$ or $\tilde{t}_1$.

Note that, while in BP5 the spectrum must include additional light sleptons or squarks in mass very close to the LSP if 
the upper bound on the relic density 
is to be satisfied, BP6 contains all the ingredients in it, as the correct 
relic abundance results from co-annihilation with the lightest stop/sbottom.\smallskip    

$\bullet$ {\it Light gluino scenario.} 
There might be finally the possibility of fitting the ATLAS signal in the presence of a 
light gluino with mass in the $\sim 1.1-1.2\tev$ range, if the stop has 
approximately the same mass of a higgsino LSP with $m_{\chi_1^0,\charone} \gsim 800\gev$.
This scenario is not excluded by the ATLAS searches described in \refsec{num}. 
Gluino pair-production provides in this case a cross section of the right magnitude, and the decay chain follows 
$\tilde{g}\to  t\,\tone$, with a 
subsequent decay of the stop to the LSP and a $c$-jet (with flavor changing loop-generated coupling). 
An exemplary spectrum is characterized by $\mgluino=1160\gev$, $\mtone=858\gev$, and $m_{\neutone}=830\gev$.

We do not report this scenario in \reftable{tab:benchmarks} and
\reffig{fig:spectra} as it seems to be excluded by CMS all-hadronic 
searches\cite{CMS-PAS-SUS-16-014,CMS-PAS-SUS-16-015,CMS-PAS-SUS-16-016} 
when the bound interpreted in the SMS T2bbbb is taken at face value. However, one must be aware that the 
acceptance/efficiency to all-hadronic final states is in this case reduced by a factor of 1.6 with respect to the T2bbbb SMS.

%%%%%%%%%%%%%%%%%%%%%%%%%%%%%%%%%%%%%%%%%%%%%%%%%%%%%%%%%%%%
\section{Approximate global analysis}\label{global}

The excess appearing in the DM-low bin of the ATLAS 1-lepton search shows a local significance greater than $3\sigma$. 
We have shown in \refsec{res} that the excess admits an explanation in the context of the MSSM for several decay chains.
On the other hand, one may wonder whether the emerging signal will maintain a noticeable significance 
with respect to the SM when other searches are considered and combined in an approximate global analysis.
We attempt to do so in this section. 

To begin with, let us notice that many of the kinematical bins constructed by the experimental collaborations show either downward- or upward fluctuations in the number of the observed events. It means that the SM (which corresponds to the background-only hypothesis) does not need to present the best fit to the observed data. To quantify to what extent the fit improves (or worsens) if a putative SUSY signal is assumed, we proceed as follows. 
For each search we  construct a test statistics as in \refeq{eq:chitot}, where again
$\Delta \chi^2_i$ are defined as $-2\log(\mathcal{L}_i/\mathcal{L}_0)$. 
In every bin $i$ the numerator, $\mathcal{L}_i$, will either assume the value corresponding to a benchmark SUSY signal,
$s_i$, or to the SM background, $s_i=0$.
The denominator, $\mathcal{L}_0$, will be instead given by the maximum likelihood value, obtained when $s_i\approx o_i-b_i$.
Note that this approach may seem to differ from the one employed in the previous sections to fit the 1-lepton signal and to derive the exclusion bounds from the other ATLAS searches, where we always normalized the likelihoood function to the background-only hypothesis. We do so for two reasons. First, we would like to have only positive test statistics. Second, it allows us to quantify how far the SM predictions are from the data.
One should also keep in mind that changes in the normalization factor result only in a shift of the $\chi^2$ value, which 
affect in equal ways the signal and background-only hypotheses.  

When combining the test statistics for individual bins, we always follow the experimental prescription. 
Thus, if for a given search the experimental collaboration explicitly states that the presented bins are independent, our test statistics will be the sum of the 
$\Delta \chi^2_i$\,; if, on the contrary, the experimental collaboration presents bins that are not independent, we construct the test statistics 
as is done in the experimental paper to derive the exclusion bounds. Depending on the search, this is given by \refeq{eq:chitot} or some modifications
of it. 
%As a rule of thumb, one should in general expect greater absolute values for the test statistics obtained from independent bins than otherwise. However,
%this should not introduce much bias, as 
The quantity of relevance to assess the goodness of fit will be the relative difference between 
the global test statistics of the SM and that of the benchmark points.

%%%%%%%%%%%%%%%%%%%%%%%%%%%%%%%%%%%%%%%%%%%%%%%%%%%%%%%
\begin{table}[t]
 \centering
 \begin{tabular}{|c|c|c|c|c|c|c|c|c|} 
 \hline
 Benchmark & SM & BP1 & BP2 & BP3 & BP4 & BP5 & BP6 \\
 \hline
 \hline
 ATLAS 1-lepton & 11.6 & 0.6 & 1.4 & 2.2 & 1.3 & 1.0 & 0.6 \\
 CMS 1-lepton & 9.8 & 11.3 & 11.0 & 10.0 & 11.1 & 8.8 & 10.2 \\
 \hline
 1-lepton $\chi^2$ & 21.4 & 11.9 & 12.4 & 12.2 & 12.4 & 9.8 & 10.8\\
 1-lepton $\sigma$ & -- & 3.1 & 3.0 & 3.0 & 3.0 & 3.4 & 3.3\\
 \hline
 \hline
 0 leptons + (b-)jets\cite{ATLAS-CONF-2016-077}& 4.6 & 1.7 & 4.2 & 1.9 & 1.4 & 4.1 & 5.6\\
 1 lepton + jets\cite{ATLAS-CONF-2016-054} & 1.4 & 3.2 & 5.6 & 3.0 & 2.1 & 4.0 & 5.3\\
 0 leptons + $2-6$ jets\cite{ATLAS-CONF-2016-078}& 3.5 & 3.2 & 3.2 & 4.4 & 2.6 & 7.1 & 5.7 \\
 $\geq 3$ b-tagged jets\cite{ATLAS-CONF-2016-052}& 5.0 & 3.1 & 6.7 & 2.9 & 2.4 & 5.0 & 4.4\\
 2 b-tagged jets\cite{Aaboud:2016nwl}& 2.2 & 2.3 & 2.4 & 2.4 & 2.5 & 2.4 & 2.5 \\
 \hline
 \hline
 Total $\chi^2_{\rm{add}}$  & 38.1 & 25.4 & 34.5 & 26.8  & 23.4 & 32.4 & 34.3 \\
 1-lepton $\sigma_{\rm{add}}$ & -- & 3.6  & 1.9  & 3.4   & 3.8  & 2.4  & 1.9\\
 \hline
 Total $\chi^2_{\rm{max}}$  & 21.4 & 11.9 & 12.4 & 12.2 & 12.4 & 9.8 & 10.8 \\
 1-lepton $\sigma_{\rm{max}}$ & -- & 3.1  & 3.0  & 3.0  & 3.0  & 3.4 & 3.3\\
 \hline
 \end{tabular}
 \caption{Breakdown of the $\chi^2$ contributions due to different searches for the background only hypothesis (SM) 
and the benchmark points.}
\label{tab:chisq}
\end{table}
%%%%%%%%%%%%%%%%%%%%%%%%%%%%%%%%%%%%%%%%%%%%%%%%%%%%%%%%%

In \reftable{tab:chisq} we show the breakdown of the $\chi^2$ values for the implemented LHC searches for the SM and for 
the postulated SUSY signals. 
In this presentation we do not consider the results from the 3.2\invfb\ analyses, 
if a higher luminosity update is available, and we also neglect searches that show a negligible $\chi^2$
contribution.
The complete $\chi^2$ breakdown for every bin of every search considered in \reftable{tab:chisq} can be found in Appendix~\ref{CMS_ATLAS_bins}.

As expected, in the ATLAS 1-lepton search there is a large 
contribution to the SM $\chi^2$ coming from the bin with the largest excess, DM-low. At the same
time, the BSM signals fit all bins to good accuracy, as was discussed in \refsec{res}.

For the corresponding analysis by CMS, the difference between the SM and the putative signals is not as marked,
and it is encouraging to see that the benchmark points are not strongly disfavored with respect to the SM even in the absence of 
a noticeable excess above the background. Note that 
contributions to the SM $\chi^2$ in the CMS search originate in some bins from an observed downward fluctuation of the background, in others from slight excesses. 
On the other hand, $\chi^2$ contributions from the bins with downward fluctuations are smoothed out by large uncertainties in the background
determination so that they do not significantly spoil the fit to the benchmark signals significantly. At the same time, the benchmark signals 
fit the small excesses better than the SM. 
The overall result is that in the CMS search the goodness of the fit is only slightly worse than in the SM, and for BP5 it is
actually better. 

To derive the total significance of the ATLAS excess after the CMS data is taken into account we combine both searches by adding the individual $\chi^2$ contributions, as the results of the two collaborations can be safely assumed to be independent.
The total significance, calculated from the chi-squared difference with the SM, is for all 6 benchmark points greater than $3\sigma$.

When we add the other ATLAS searches to the global fit, the significance of the BSM signal with respect to the SM predictions 
can either improve or worsen, depending on the particular benchmark model.
For the 0 lepton + b-jets search\cite{ATLAS-CONF-2016-077} almost all the models describe the experimental results better than 
the background-only hypothesis. That is once more due to several 
small excesses observed in this search that are better fitted by the putative SUSY signals than by the SM background.
This could be a hint that the 1-lepton excess is a real phenomenon, but most likely the good fit results from the fact that both 
searches are optimized for stop production and the background determination is to some degree correlated in one and the other search.

All benchmark models show instead some tension with 
the 1 lepton + jets analysis\cite{ATLAS-CONF-2016-054}. The tension originiates from one bin, optimized for squark pair production, in which no excess is observed. Kinematical cuts employed in this bin are very similiar to those in the bin DM-low of the ATLAS 1-lepton search, so large signals for our benchmark models are to be expected. The tension is the strongest for the scenarios BP2 and BP6, where it reaches approximately the $2\sigma$ level. Note that these are the models that predict the largest signals in the DM-low bin.
Additionally, models BP5 and BP6 are disfavored at 1.9 and $1.5\,\sigma$, respectively, by the all hadronic search\cite{ATLAS-CONF-2016-078}. That is also to be expected, as both of them present relatively light left-chiral squarks 
for which this search is optimized. 
%Finally, model BP2 shows $1.4\sigma$ tension with the 3 b-tagged jets search\cite{ATLAS-CONF-2016-052}. The signal in this case is enhanced by jets coming from decays of gluino  with the mass of 1900 GeV,  much lighter than for other benchmarks scenarios.

%large $\chi^2$ contribution for both the SUSY signals and the SM in the 3 b-tagged search\cite{ATLAS-CONF-2016-052} originiates from various sources. In the SM case it comes entirely from two bins, one of which  presents a $2\sigma$ excess and the other $1.6\sigma$ deficit in the number of the observed events. In the case of the signal three b-jets come from decays of gluino, which have mass of 1900 GeV. BP2 shows $1.4\sigma$ tension due to ...\kk{End}

In the two bottom rows  we present the total $\chi^2$ for all the searches listed in \reftable{tab:chisq}, 
and the resulting total significance of the ATLAS 1-lepton excess. When combining the results of different analyses, we use two approaches. 
First, we assume that all ATLAS searches are statistically independent and we simply sum the corresponding $\chi^2$ contributions. 
We mark the results thus obtained with the subscript ``add.''  
As a result, the significance of the 1-lepton excess decreases for BP2, BP5, and BP6.

As we have mentioned above, however, from the outside it is hard to gauge to what extent ATLAS searches, 
even the ones with different final states topologies, can be considered independent from one another. 
The ATLAS Collaboration itself, in their global analysis of 20 searches based on the 8\tev\ data\cite{Aad:2015baa}, 
decided to employ for overall exclusion limits a so called ``best-of'' strategy, which only uses the result of the analysis with the best sensitivity.
To mimic the same approach, we also provide here significances based on the maximum $\chi^2$ only, dubbed here as ``max.''
Since we are not able to quantify the correlations between different ATLAS searches, we think it is safe to assume that the real 
significance lies somewhere in between the ``add'' and ``max'' scenarios.

\bigskip
One might wonder if the scenarios described above 
give rise to specific signatures at the LHC (besides a clear 1-lepton signal) that might help distinguish them from one another.
Since all scenarios involve the production of sparticles with color charge, 0-lepton searches with different kinematical variables should 
produce a complementary, if weaker, signal when the integrated luminosity is increased by 
a factor $\sim 3$ or more with respect to the amount of data presented at ICHEP.

In particular, by taking the 0-lepton ATLAS search\cite{ATLAS-CONF-2016-078} as an example, 
and rescaling there the SM background to $\sim50\invfb$, 
our numerical simulation shows bins with clear excesses above the background for BP5, BP6, and possibly BP3, where we have 
ordered the benchmarks by decreasing significance. Whereas, with the same luminosity, BP1,  BP2, and BP4 would still be indistinguishable from the background in 
a search like\cite{ATLAS-CONF-2016-078}.

With $\sim 100\invfb$ BP5 and BP6 will probably produce a $>5\,\sigma$ discovery in the 0-lepton search, 
and significant excesses would appear for BP3 and BP2, in decreasing order of signal strength.
Note, however, that even at this high luminosity, BP1 will remain difficult to observe in searches with a hard lepton veto, 
because of its large branching ratio to the 1-lepton final state, which allows for a clear fitting of our signal in the first place.
The reader should also keep in mind that, at least for the data provided at $\sim 13\invfb$, the corresponding CMS analyses have shown 
better sensitivity in the parameter space characterized by a moderately heavy LSP so that the benchmark points 
should appear earlier and to greater significance in a search like the CMS $\alpha_T$.

Finding appropriate UV completions to the scenarios presented above is beyond the purpose 
of this paper, although it might become a necessary endeavour were the excesses to be confirmed in the next round of data.
Models with light stops in general do not fare well with boundary conditions defined at the GUT scale, 
because the current bound on the gluino mass ($\mgluino\geq 1.6-1.8\tev$ for the most common SUSY spectra) 
translates into a large renormalization of the stop mass at the low scale, independently of the initial choice for GUT-scale boundary conditions.

Gauge mediation remains a viable option, as it allows some freedom in the choice of the messenger scale, 
and in particular models with matter-messenger mixing like the one proposed in, e.g.,\cite{Chacko:2001km} and several other papers, 
can easily produce right-chiral stops and binos in agreement with the spectrum of BP1. 
The remaining benchmark points, however, present more involved and often compressed spectra, which might prove more challenging 
from the model-building point of view.

%%%%%%%%%%%%%%%%%%%%%%%%%%%%%%%%%%%%%%%%%%%%%%%%%%%%%%%%%%%%
\section{Summary and conclusions}\label{sum}

In this paper we have investigated the possibility that some mild excesses 
over the SM background that emerged in the bins of the ATLAS~1-lepton + (b-)jets + $E_T^{\textrm{miss}}$ 
search reported at the ICHEP 2016 conference are due to the first appearance 
of supersymmetry. Working in the framework of the p19MSSM, we have determined a few different types of SUSY spectra 
than can explain the observed data and are not excluded by other direct searches for SUSY.

The most straightforward possibility (BP1 in \reftable{tab:benchmarks} and \reffig{fig:spectra}) corresponds to the simplified model spectrum 
used by the ATLAS Collaboration for the interpretation of the search results, which can fit the putative signal with light right-chiral stops of 
about 600-800\gev\ and a bino-like neutralino LSP with a mass of the order of $\sim 400\gev$. 
One of the advantages of this scenario, beyond its simplicity, is that it can be relatively easily embedded 
in known UV completions, for example 
models of gauge mediation with matter-messenger mixing terms in the superpotential that allow for light right-handed stops.

The signal can also be fitted if charginos of mass in between the stop and the neutralino are present in the spectrum (BP2),
where in this case the stop can be slightly lighter than in BP1,
as a light chargino has the effect of reducing the branching ratio to the 1-lepton final state topology.
Additional good fits can be achieved when the signal is produced by left chiral stop/sbottoms, 
because of the boost in production cross section that arises by including the $\tilde{b}_1\to t\charone$ topology (BP3), and when
the right-chiral sbottom is also added to the light spectrum (BP4).

For an LSP of mass significantly above $400-500\gev$ 
the acceptance for the signal produced by relatively light stops and sbottoms drops drastically. 
More complex spectra, however, involving associated production of light-generation squarks and electroweakinos 
with a mass of about $1.1\tev$ and cascade decays to stops and sbottoms (BP5, BP6) can still
produce the right signal even for a $\sim700\gev$ LSP, 
because of the increase in production cross section due to the presence of four degenerate ``valence'' squarks.
These light valence squarks are in pole position to be probed by all-hadronic searches in the next round of the LHC data.

For all of the presented benchmark-point scenarios we have compared their global $\chi^2$ with respect to the ATLAS and CMS searches 
listed in \reftable{tab:chisq} with the corresponding $\chi^2$ of the background-only hypothesis. Our computation shows that
the signal hypothesis is, at least slightly, favored globally with respect to the SM, and this is true under different choice of test statistics.

On the other hand, we conclude by repeating once again that the analysis presented here is meant to investigate a possibility that might reveal 
itself as a statistical fluctuation. On the positive side, if the signal is real it will be confirmed with more data and the signatures described in \refsec{res} will be useful to distinguish one or the other scenario.

%%%%%%%%%%%%%%%%%%%%%%%%%%%%%%%%%%%%%%%%%%%%%%%%%%%%%%%%%%%%%%%%%%%%%%%%%%%%%%%%
\bigskip  
\begin{center}
\textbf{ACKNOWLEDGMENTS}
\end{center}
We thank the anonymous Referee for suggesting the global analysis of \refsec{global}.
K.K. is supported in part by the DFG Research Unit FOR 1873 
``Quark Flavour Physics and Effective Field Theories''. 
The use of the CIS computer cluster at the National Centre for Nuclear Research in Warsaw is gratefully acknowledged. 
\bigskip
%%%%%%%%%%%%%%%%%%%%%%%%%%%%%%%%%%%%%%%%%%%%%%%%%%%%%%
\appendix

\section{Validation of the CMS 1 lepton search}\label{CMS_valid}

\begin{figure}[t]
\centering
\includegraphics[width=0.50\textwidth]{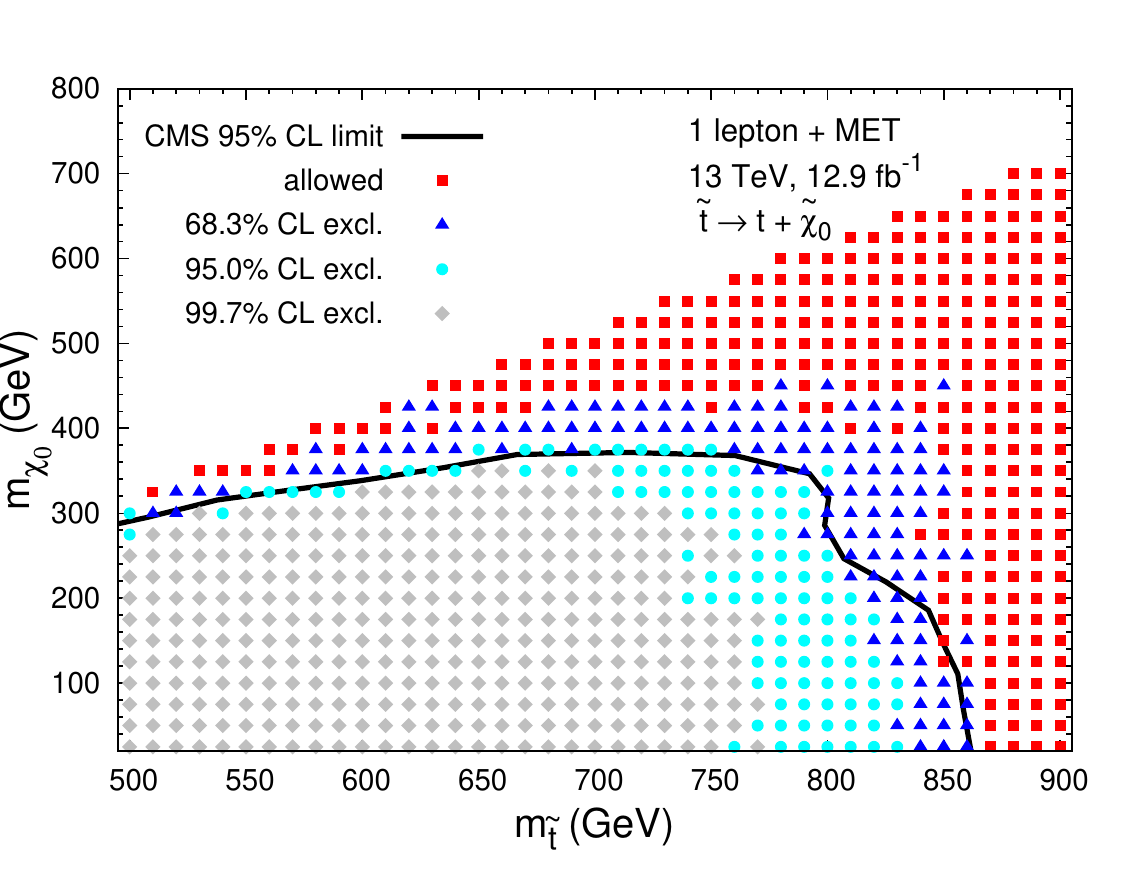}
\caption[]{ Our simulation of the CMS 1-lepton search with 12.9 \invfb\ of data for direct stop production and a decay $\tilde{t}_1\to t \neutone$. Points that are excluded at the 99.7\%~C.L. are shown as gray diamonds, at the 95.0\%~C.L. as cyan circles, and at the 68.3\%~C.L. as blue triangles. The points shown as red squares are considered as allowed. The solid black lines show the published 95\%~C.L. contours by CMS.}
\label{fig:CMS_1lepton}
\end{figure}

In this appendix we provide the validation of our numerical simulation of the CMS search for direct top squark pair 
production in the single-lepton final state\cite{CMS-PAS-SUS-16-028}. 
Validation of all other searches used in this paper and listed in \refsec{num} can be found in Appendix~A
of Ref.\cite{Kowalska:2016ent}.

We check the consistency of our statistical analysis with the official bounds provided by CMS by comparing the 95\%~C.L. 
exclusion line, derived by the experimental collaboration in one of the SMS used for their result interpretation,
with the line obtained with our code in the same SMS.
The technical details of the experimental analysis can be found in\cite{CMS-PAS-SUS-16-028}. 

%Note that in the validation analyses we used the NLO+NNL cross-sections, provided by LHC SUSY Cross Section Working Group\cite{LHCSXSECWG}.

The SMS analyzed here consists of a stop decaying into top quark and neutralino LSP.
The experimental signature includes one lepton, jets and large amount of missing energy. The analysis is performed with the following pre-selection cuts:
\begin{itemize}
\item exactly one signal lepton with $p_T>20$ GeV and $|\eta|<2.4(1.44)$ for muons (electrons),
\item at least 2 signal jets with $p_T>30$ GeV and $|\eta|<2.4$,
\item at least 1 signal b-jet,
\item missing energy $\met > 250\gev$.
\end{itemize}

The kinematical variables used to discriminate between the signal and the background are: 
azimuthal angle $\dfi$ between two leading jets and missing energy; 
the transverse mass $m_T$ of the signal lepton and missing transverse momentum; 
the magnitude of the negative vectorial sum of the transverse momentum of jets with $p_T>20$ in the event, $H_T^{\rm{miss}}$, 
and the variable $M^{\rm{W}}_{\rm{T}2}$ defined in\cite{Bai:2012gs}. 
To implement the latter in the recast tool, we used the C++ code provided in\cite{Bai:2012gs}.
Fifteen exclusive signal regions are defined with various jet multiplicities and binned in $\met$ and $M^{\rm{W}}_{\rm{T}2}$.

We present in \reffig{fig:CMS_1lepton} the validation of our simulation in terms of the exclusion limits in the parameter space 
$(\mtone,m_{\neutone})$ for the direct stop production scenario and a subsequent decay into top quark and neutralino LSP.
Gray diamonds represent the points excluded by our likelihood function at the 99.7\%~C.L., 
cyan circles are excluded at the 95.0\%~C.L., 
and blue triangles are excluded at the 68.3\%~C.L. 
The solid black line shows the 95\%~C.L. CMS exclusion limit, which we overlap for comparison.

\section{Summary of all $\mathbf{\Delta\chi^2}$ contributions}\label{CMS_ATLAS_bins}

In Tables~\ref{tab:chisq_CMS}-\ref{tab:chisq_ATLAS_2bj} we present
the complete $\chi^2$ breakdown for every bin of every search considered in \reftable{tab:chisq}.
The $s_i$ column shows the number of signal events simulated according to the procedure of \refsec{num}.
The total $\Delta\chi^2$ for each search is calculated following the prescription given in the experimental paper 
and summarized in the corresponding caption.

\setlength\tabcolsep{3pt}
%%%%%%%%%%%%%%%%%%%%%%%%%%%%%%%%%%%%%%%%%%%%%%%%%%%%%%%
\begin{table}[H]\footnotesize
 \centering
 \begin{tabular}{|c|c|c||c|cc|cc|cc|cc|cc|cc|} 
 \hline
 Bin & Obs. & Bkg. & SM & \multicolumn{2}{c|}{BP1} & \multicolumn{2}{c|}{BP2} & \multicolumn{2}{c|}{BP3} & \multicolumn{2}{c|}{BP4} & \multicolumn{2}{c|}{BP5} & \multicolumn{2}{c|}{BP6} \\
 & & & $\Delta\chi^2_i$ & $s_i$ & $\Delta\chi^2_i$ & $s_i$ & $\Delta\chi^2_i$ & $s_i$ & $\Delta\chi^2_i$ & $s_i$ & $\Delta\chi^2_i$ & $s_i$ & $\Delta\chi^2_i$ & $s_i$ & $\Delta\chi^2_i$\\
 \hline
 \hline
 2j-250 & 72 & $58.6\pm 6.0$ &  1.9 & 0.6 & 1.7 & 0.9 & 1.6 & 1.5 & 1.5 & 0.8 & 1.6 & 0.4 & 1.7 & 0.2 & 1.8\\
 2j-350 & 7 & $12.2\pm 1.9$ & 1.8 & 0.3 & 2.0 & 0.1 & 1.9 & 0.2 & 2.0 & 0.4 & 2.1 & 0.1 & 1.9 & 0 & 1.8 \\
 2j-450 & 5 & $3.7\pm 0.8$ &  0.4 & 0.1 & 0.4 & 0.1 & 0.4 & 0 & 0.4 & 0.1 & 0.4 & 0 & 0.4 & 0.1 & 0.4\\
 3j-250 & 35 & $38.0\pm 4.6$ &  0.1 & 1.5 & 0.3 & 2.0 & 0.4 & 1.1 & 0.2 & 0.1 & 0.1 & 0.5 & 0.2 & 0.3 & 0.1 \\
 3j-350 & 9 & $10.4\pm 1.6$ &  0.1 & 0.5 & 0.2 & 0.8 & 0.3 & 0.2 & 0.2 & 0.7 & 0.3 & 0.3 & 0.2 & 0.3 & 0.2\\
 3j-450 & 6 & $5.3 \pm 1.1$ & 0.1 & 0.1 & 0.1 & 0.4 & 0 & 0.1 & 0.1 & 0 & 0.1 & 0 & 0.1 & 0 & 0.1\\
 3j-550 & 3 & $3.0 \pm 0.7$  & 0 & 0& 0 & 0 & 0 & 0.1 & 0 & 0 & 0 & 0.1 & 0 & 0.3 & 0\\
 4j-250-low & 121 & $141\pm 15$ & 1.1 & 2.0 & 1.3 & 1.8 & 1.3 & 2.9 & 1.4 & 2.2 & 1.3 & 1.5 & 1.2 & 1.5 & 1.2 \\
 4j-350-low & 22 & $26.6\pm 4.0$ & 0.4 & 0.9 & 0.6 & 0.5 & 0.6 & 0.8 & 0.6 & 0.7 & 0.6 & 0.7 & 0.6 & 0.4 & 0.5\\
 4j-450-low & 9 & $6.6\pm 1.6$ & 0.7 & 0.2 & 0.6 & 0 & 0.7 & 0.4 & 0.5 & 0.4 & 0.5 & 0.7 & 0.4 & 0.3 & 0.5\\
 4j-250-high & 44 & $43.3\pm 6.2$  & 0 & 5.1 & 0.2 & 5.6 & 0.2 & 4.4 & 0.1 & 5.1 & 0.2 & 2.3 & 0 & 4.8 & 0.2\\
 4j-350-high & 11 & $14.2 \pm 2.5$ & 0.4 & 3.2 & 1.8 & 3.2 & 1.8 & 1.5 & 1.0 & 3.3 & 1.9 & 1.9 & 1.2 & 3.7 & 2.1\\
 4j-450-high & 5 & $4.8\pm 1.3$  & 0 & 1.3 & 0.1 & 1.1 & 0.1 & 0.7 & 0 & 1.5 & 0.2 & 1.0 & 0 & 1.1 & 0.1 \\
 4j-550-high & 1 & $1.7 \pm 0.6$  & 0.1 & 0.6 & 0.5 & 0.4 & 0.4 & 0.5 & 0.5 & 0.5 & 0.5 & 0.1 & 0.1 & 0.3 & 0.3 \\
 4j-650-high & 3 & $0.92\pm 0.33$ & 2.7 & 0.4 & 1.5 & 0.5 & 1.3 & 0.4 & 1.5 & 0.5 & 1.3 & 0.8 & 0.8 & 0.7 & 0.9\\
 \hline
  \hline
 \multicolumn{3}{|c||}{Total $\Delta\chi^2$}& 9.8 &  \multicolumn{2}{|c|}{11.3} & \multicolumn{2}{|c|}{11.0} & \multicolumn{2}{|c|}{10.0} & \multicolumn{2}{|c|}{11.1} & \multicolumn{2}{|c|}{8.8} & \multicolumn{2}{|c|}{10.2} \\
\hline
 \end{tabular}
 \caption{Breakdown of the $\chi^2$ contributions due to different signal bins of the CMS 1-lepton search\cite{CMS-PAS-SUS-16-028} for the background-only hypothesis (SM) and the benchmark points. The total $\Delta\chi^2$ is given by the sum of each individual contribution. The collaboration has not labeled the bins by name. Nevertheless, we do provide an indicative label in column 1, based on the number of jets and the $\met$ cut adopted. }
\label{tab:chisq_CMS}
\end{table}
%%%%%%%%%%%%%%%%%%%%%%%%%%%%%%%%%%%%%%%%%%%%%%%%%%%%%%%%%

%%%%%%%%%%%%%%%%%%%%%%%%%%%%%%%%%%%%%%%%%%%%%%%%%%%%%%%
\begin{table}[H]\footnotesize
 \centering
 \begin{tabular}{|c|c|c||c|cc|cc|cc|cc|cc|cc|} 
 \hline
 Bin & Obs. & Bkg. & SM & \multicolumn{2}{c|}{BP1} & \multicolumn{2}{c|}{BP2} & \multicolumn{2}{c|}{BP3} & \multicolumn{2}{c|}{BP4} & \multicolumn{2}{c|}{BP5} & \multicolumn{2}{c|}{BP6} \\
& & &  $\Delta\chi^2_i$ & $s_i$ & $\Delta\chi^2_i$ & $s_i$ & $\Delta\chi^2_i$ & $s_i$ & $\Delta\chi^2_i$ & $s_i$ & $\Delta\chi^2_i$ & $s_i$ & $\Delta\chi^2_i$ & $s_i$ & $\Delta\chi^2_i$\\
 \hline
 \hline
SR1 & 37 & $24 \pm 3$ & 4.5 & 11.4 & 0.1 & 9.8 & 0.3 & 8.3 & 0.6 & 8.6 & 0.5 & 9.9 & 0.2 & 10.9 & 0.1\\
tN\_high & 5 & $3.8 \pm 0.8$  & 0.4 & 3.2 & 0.5 & 4.5 & 1.4 & 2.4 & 0.2 & 3.0 & 0.4 & 4.0 & 1.0 & 3.3 & 0.6 \\
bC2x\_diag & 37 & $22 \pm 3$ & 6.1 & 11.1 & 0.4 & 12.1 & 0.2 & 9.7 & 0.7 & 7.8 & 1.3 & 9.0 & 0.9 & 11.7 & 0.3 \\
bC2x\_med & 14 & $13 \pm 2$ & 0.1 & 2.7 & 0.1 & 4.0 & 0.4 & 2.4 & 0.1 & 2.3 & 0.1 & 1.7 & 0 & 4.0 & 0.4\\
bCbv & 7 & $7.4 \pm 1.8$ & 0 & 1.0 & 0.1 & 0.6 & 0 & 0.6 & 0 & 0.9 & 0.1 & 1.6 & 0.3 & 1.6 & 0.3 \\
DM\_low & 35 & $17 \pm 2$ & 11.6 & 13.6 & 0.6 & 14.4 & 0.4 & 9.4 & 2.2 & 11.9 & 1.1 & 12.5 & 0.9 & 13.9 & 0.5 \\
DM\_high & 21 & $15 \pm 2$ & 1.8 & 6.1 & 0 & 7.4 & 0.1 & 4.4 & 0.1 & 5.4 & 0 & 6.0 & 0 & 6.7 & 0\\
 \hline
   \hline
 \multicolumn{3}{|c||}{Total $\Delta\chi^2$}& 11.6 &\multicolumn{2}{|c|}{0.6} & \multicolumn{2}{|c|}{1.4} & \multicolumn{2}{|c|}{2.2} & \multicolumn{2}{|c|}{1.3} & \multicolumn{2}{|c|}{1.0} & \multicolumn{2}{|c|}{0.6}  \\
\hline
 \end{tabular}
 \caption{Breakdown of the $\chi^2$ contributions due to different signal bins of the ATLAS 1-lepton search\cite{ATLAS-CONF-2016-050} for the background-only hypothesis (SM) and the benchmark points. The total $\Delta\chi^2$ is given by the maximum of the individual contributions. The bins are labeled according to their ATLAS name. }
\label{tab:chisq_ATLAS_1l}
\end{table}
%%%%%%%%%%%%%%%%%%%%%%%%%%%%%%%%%%%%%%%%%%%%%%%%%%%%%%%%%

%%%%%%%%%%%%%%%%%%%%%%%%%%%%%%%%%%%%%%%%%%%%%%%%%%%%%%%
\begin{table}[H]\footnotesize
 \centering
 \begin{tabular}{|c|c|c||c|cc|cc|cc|cc|cc|cc|} 
 \hline
Bin & Obs. & Bkg. & SM & \multicolumn{2}{c|}{BP1} & \multicolumn{2}{c|}{BP2} & \multicolumn{2}{c|}{BP3} & \multicolumn{2}{c|}{BP4} & \multicolumn{2}{c|}{BP5} & \multicolumn{2}{c|}{BP6} \\
& & & $\Delta\chi^2_i$ & $s_i$ & $\Delta\chi^2_i$ & $s_i$ & $\Delta\chi^2_i$ & $s_i$ & $\Delta\chi^2_i$ & $s_i$ & $\Delta\chi^2_i$ & $s_i$ & $\Delta\chi^2_i$ & $s_i$ & $\Delta\chi^2_i$\\
 \hline
 \hline
Meff\_2j\_0800 & 650 & $610 \pm 50$ & 0.5 & 5.7 & 0.4 & 8.4 & 0.3 & 6.7 & 0.4 & 8.4 & 0.3 & 17.1 & 0.2 & 20.0 & 0.1 \\
Meff\_2j\_1200& 270 & $297\pm 29$ & 0.6 & 5.3 & 0.9 & 5.9 & 0.9 & 7.7 & 1.0 & 4.7 & 0.9 & 16.3 & 1.6 & 15.1 & 1.5\\
Meff\_2j\_1600& 96 & $121\pm 13$ & 2.2 & 3.9 & 2.9 & 4.5 & 3.1 & 6.5 & 3.5 & 2.0 & 2.6 & 13.7 & 5.3 & 10.9 & 4.5 \\
Meff\_2j\_2000& 29 & $42\pm 6$ & 2.2 & 2.6 & 3.2 & 2.6 & 3.2 & 5.2 & 4.4 & 1.1 & 2.6 & 10.3 & 7.1 & 7.8 & 5.7\\
Meff\_3j\_1200& 363 & $355\pm 33$ & 0.1 & 7.7 & 0 & 11.2 & 0 & 17.1 & 0 & 7.8 & 0 & 42.9 & 0.8 & 31.5 & 0.4\\
Meff\_4j\_1000& 97 & $84\pm 7$ & 1.3 & 4.1 & 0.6 & 5.5 & 0.4 & 3.4 & 0.7 & 4.6 & 0.5 & 7.3 & 0.3 & 13.6 & 0\\
Meff\_4j\_1400& 71 & $66\pm 8$ & 0.2 & 6.0 & 0 & 7.4 & 0 & 5.1 & 0 & 6.2 & 0 & 8.8 & 0.1 & 15.9 & 0.8\\
Meff\_4j\_1800& 37 & $27.0\pm 3.2$ & 2.5 & 3 & 1.2 & 5.3 & 0.5 & 4.3 & 0.8 & 3.0 & 1.2 & 6.8 & 0.3 & 9.3 & 0\\
Meff\_4j\_2200& 10 & $4.8\pm 1.1$ & 3.5 & 0.9 & 2.3 & 2.0 & 1.2 & 1.4 & 1.7 & 0.9 & 2.3 & 1.5 & 1.6 & 2.6 & 0.8\\
Meff\_4j\_2600& 3 & $2.7\pm 0.6$ & 0.1 & 0.7 & 0 & 1.9 & 0.5 & 0.8 & 0 & 0.4 & 0 & 1.0 & 0.1 & 1.6 & 0.4\\
Meff\_5j\_1400& 64 & $68\pm 9$ & 0.1 & 4.7 & 0.4 & 5.4 & 0.5 & 5.9 & 0.6 & 3.6 & 0.3 & 14.7 & 2.2 & 12.1 & 1.6\\
Meff\_6j\_1800& 10 & $5.5\pm 1.0$ & 2.6 & 2.3 & 0.6 & 3.7 & 0.1 & 2.5 & 0.5 & 2.0 & 0.7 & 5.3 & 0 & 6.1 & 0.2\\
Meff\_6j\_2200& 1 & $0.82\pm 0.35$ & 0.1 & 0.2 & 0 & 0.7 & 0.1 & 0.4 & 0 & 0.2 & 0 & 0.6 & 0.1 & 0.8 & 0.2\\
 \hline
    \hline
 \multicolumn{3}{|c||}{Total $\Delta\chi^2$}& 3.5 &  \multicolumn{2}{|c|}{3.2} & \multicolumn{2}{|c|}{3.2} & \multicolumn{2}{|c|}{4.4} & \multicolumn{2}{|c|}{2.6} & \multicolumn{2}{|c|}{7.1} &\multicolumn{2}{|c|}{5.7} \\
\hline
 \end{tabular}
 \caption{Breakdown of the $\chi^2$ contributions due to different signal bins of the ATLAS 0 leptons + $2-6$ jets search\cite{ATLAS-CONF-2016-078} for the background-only hypothesis (SM) and the benchmark points. The total $\Delta\chi^2$ is given by the maximum of the individual contributions. The bins are labeled according to their ATLAS name. }
\label{tab:chisq_ATLAS_26j}
\end{table}
%%%%%%%%%%%%%%%%%%%%%%%%%%%%%%%%%%%%%%%%%%%%%%%%%%%%%%%%%

%%%%%%%%%%%%%%%%%%%%%%%%%%%%%%%%%%%%%%%%%%%%%%%%%%%%%%%
\begin{table}[H]\footnotesize
%\vspace*{1.6cm}
 \centering
 \begin{tabular}{|c|c|c||c|cc|cc|cc|cc|cc|cc|} 
 \hline
 Bin & Obs. & Bkg. & SM & \multicolumn{2}{c|}{BP1} & \multicolumn{2}{c|}{BP2} & \multicolumn{2}{c|}{BP3} & \multicolumn{2}{c|}{BP4} & \multicolumn{2}{c|}{BP5} & \multicolumn{2}{c|}{BP6} \\
& & & $\Delta\chi^2_i$ & $s_i$ & $\Delta\chi^2_i$ & $s_i$ & $\Delta\chi^2_i$ & $s_i$ & $\Delta\chi^2_i$ & $s_i$ & $\Delta\chi^2_i$ & $s_i$ & $\Delta\chi^2_i$ & $s_i$ & $\Delta\chi^2_i$\\
 \hline
 \hline
SRA-TT & 8 & $5.2\pm 1.4$ & 1.1 & 4.0 & 0.1 & 9.4 & 3.1 & 3.4 & 0 & 5.6 & 0.6 & 8.5 & 2.4 & 9.9 & 3.5  \\
SRA-TW & 5 & $5.7 \pm 1.6$ & 0 & 1.9 & 0.6 & 2.7 & 1.1 & 1.7 & 0.5 & 2.0 & 0.7 & 3.6 & 1.7 & 3.9 & 1.9\\
SRA-T0 & 16 & $11.3 \pm 2.6$ & 1.2 & 2.2 & 0.4 & 4.4 & 0 & 2.5 & 0.3 & 3.9 & 0.1 & 4.5 & 0 & 7.2 & 0.2\\
SRB-TT & 17 & $10.6\pm 2.3$ & 2.3 & 4.4 & 0.2 & 8.5 & 0.1 & 3.3 & 0.5 & 6.2 & 0 & 3.9 & 0.4 & 10.4 & 0.6\\
SRB-TW & 18 & $16.7\pm 3.6$ & 0.1 & 3.6 & 0.1 & 4.8 & 0.3 & 2.5 & 0 & 4.0 & 0.2 & 3.5 & 0.1 & 5.7 & 0.5 \\
SRB-T0 & 84 & $60 \pm 14$ & 2.2 & 5.3 & 1.4 & 7.6 & 1.1 & 4.8 & 1.4 & 9.3 & 0.9 & 3.8 & 1.6 & 10.1 & 0.8 \\
SRC-low & 36 & $23.9 \pm 7.5$ & 1.8 & 0.7 & 1.6 & 1.9 & 1.3 & 0.7 & 1.6 & 1.4 & 1.4 & 0.3 & 1.7 & 1.5 & 1.4\\
SRC-med & 14 & $9.4 \pm 3.5$ & 1.0 & 0.3 & 0.9 & 1.3 & 0.6 & 0.3 & 0.9 & 0.7 & 0.8 & 0 & 1.0 & 0.8 & 0.7\\
SRC-high & 9 & $10.5\pm 3.7$ & 0 & 0.5 & 0.1 & 1.2 & 0.2 & 0.1 & 0 & 1.1 & 0.2 & 0.2 & 0.1 & 1.1 & 0.2 \\
SRE & 9 & $7.1\pm 1.8$ & 0.4 & 3.4 & 0.1 & 5.9 & 1.0 & 3.2 & 0.1 & 4.4 & 0.4 & 3.7 & 0.2 & 5.4 & 0.8\\
SRF & 3 & $2.8\pm 1.0$ & 0.1 & 2.0 & 0.5 & 3.6 & 1.8 & 1.8 & 0.4 & 2.0 & 0.5 & 5.4 & 3.7 & 5.2 & 3.5\\
 \hline
     \hline
 \multicolumn{3}{|c||}{Total $\Delta\chi^2$}& 4.6 &  \multicolumn{2}{|c|}{1.7} & \multicolumn{2}{|c|}{4.2} & \multicolumn{2}{|c|}{1.9} & \multicolumn{2}{|c|}{1.4} & \multicolumn{2}{|c|}{4.1} &\multicolumn{2}{|c|}{5.6} \\
\hline
 \end{tabular}
 \caption{Breakdown of the $\chi^2$ contributions due to different signal bins of the ATLAS 0 leptons + (b-)jets search\cite{ATLAS-CONF-2016-077} for the background-only hypothesis (SM) and the benchmark points. 
Bins 1, 2, and 3 are orthogonal to each other, and so are bins 4, 5, and 6.
The total $\Delta\chi^2$ is given by $\max\{\sum_{i=1}^3\Delta\chi^2_i,\sum_{i=4}^6\Delta\chi^2_i,\Delta\chi^2_7,\Delta\chi^2_8,\Delta\chi^2_9,\Delta\chi^2_{10},\Delta\chi^2_{11}\}$. The bins are labeled according to their ATLAS name. }
\label{tab:chisq_ATLAS_0l}
\end{table}
%%%%%%%%%%%%%%%%%%%%%%%%%%%%%%%%%%%%%%%%%%%%%%%%%%%%%%%%%

%%%%%%%%%%%%%%%%%%%%%%%%%%%%%%%%%%%%%%%%%%%%%%%%%%%%%%%
\begin{table}[H]\footnotesize
%\vspace*{-0.8cm}
 \centering
 \begin{tabular}{|c|c|c||c|cc|cc|cc|cc|cc|cc|} 
 \hline
Bin &  Obs. & Bkg. & SM & \multicolumn{2}{c|}{BP1} & \multicolumn{2}{c|}{BP2} & \multicolumn{2}{c|}{BP3} & \multicolumn{2}{c|}{BP4} & \multicolumn{2}{c|}{BP5} & \multicolumn{2}{c|}{BP6} \\
& & & $\Delta\chi^2_i$ & $s_i$ & $\Delta\chi^2_i$ & $s_i$ & $\Delta\chi^2_i$ & $s_i$ & $\Delta\chi^2_i$ & $s_i$ & $\Delta\chi^2_i$ & $s_i$ & $\Delta\chi^2_i$ & $s_i$ & $\Delta\chi^2_i$\\
 \hline
 \hline
SR-Gbb-A & 2 & $1.6\pm 0.7$ & 0.2 & 0.3 & 0 & 3.9 & 2.6 & 0.4 & 0 & 0.6 & 0 & 0.9 & 0 & 1.3 & 0.2 \\
SR-Gbb-B & 15 & $21\pm 5$ & 0.5 & 2.0 & 1.1 & 4.0 & 1.9 & 0.3 & 0.6 & 1.6 & 1.0 & 3.6 & 1.7 & 4.2 & 2.0 \\
SR-Gtt-0L-A & 1 & $0.94\pm 0.31$ & 0 & 0.6 & 0.2 & 3.3 & 3.4 & 0.7 & 0.2 & 0.9 & 0.4 & 2.4 & 2.1 & 2.1 & 1.7 \\
SR-Gtt-0L-B & 11 & $5.0 \pm 1.5$ & 3.9 & 2.3 & 1.4 & 4.6 & 0.2 & 1.8 & 1.8 & 2.9 & 0.9 & 5.6 & 0 & 5.2 & 0.1 \\
SR-Gtt-1L-A & 1 & $1.0 \pm 0.6$ & 0.2 & 0 & 0.2 & 1.4 & 0.7 & 0 & 0.2 & 0.1 & 0.1 & 0.4 & 0 & 0.2 & 0 \\
SR-Gtt-1L-B & 2 & $1.1 \pm 0.4$ & 0.6 & 0.2 & 0.4 & 1.2 & 0 & 0.1 & 0.5 & 0.1 & 0.5 & 0.5 & 0.1 & 0.2 & 0.4 \\
SR-Gtt-1L-C & 4 & $7.0 \pm 2.8$& 0.4 & 0.4 & 0.6 & 0.4 & 0.6 & 0.1 & 0.4 & 0.3 & 0.5 & 1.6 & 1.2 & 0.6 & 0.7\\
\hline
    \hline
 \multicolumn{3}{|c||}{Total $\Delta\chi^2$}& 5.0 &  \multicolumn{2}{|c|}{3.1} & \multicolumn{2}{|c|}{6.7} & \multicolumn{2}{|c|}{2.9} & \multicolumn{2}{|c|}{2.4} & \multicolumn{2}{|c|}{5.0} &\multicolumn{2}{|c|}{4.4} \\
\hline
 \end{tabular}
 \caption{Breakdown of the $\chi^2$ contributions due to different signal bins of the ATLAS  3 b-tagged jets search\cite{ATLAS-CONF-2016-052} for the background-only hypothesis (SM) and the benchmark points. For the first two bins the largest $\Delta\chi^2_i$ is chosen. The chi-squared of bins 3 and 4 are instead confronted in a pairwise manner with the chi-squared of bins 5, 6, and 7, and the combination with the largest contribution is selected. The total $\Delta\chi^2$ is given by the sum of the two partial chi-squared. The bins are labeled according to their ATLAS name. 
 }
\label{tab:chisq_ATLAS_3bj}
\end{table}
%%%%%%%%%%%%%%%%%%%%%%%%%%%%%%%%%%%%%%%%%%%%%%%%%%%%%%%%%

%%%%%%%%%%%%%%%%%%%%%%%%%%%%%%%%%%%%%%%%%%%%%%%%%%%%%%%
\begin{table}[H]\footnotesize
\vspace*{0.1cm}
 \centering
 \begin{tabular}{|c|c|c||c|cc|cc|cc|cc|cc|cc|} 
 \hline
Bin & Obs. & Bkg. & SM & \multicolumn{2}{c|}{BP1} & \multicolumn{2}{c|}{BP2} & \multicolumn{2}{c|}{BP3} & \multicolumn{2}{c|}{BP4} & \multicolumn{2}{c|}{BP5} & \multicolumn{2}{c|}{BP6} \\
& & & $\Delta\chi^2_i$ & $s_i$ & $\Delta\chi^2_i$ & $s_i$ & $\Delta\chi^2_i$ & $s_i$ & $\Delta\chi^2_i$ & $s_i$ & $\Delta\chi^2_i$ & $s_i$ & $\Delta\chi^2_i$ & $s_i$ & $\Delta\chi^2_i$\\
 \hline
 \hline
GG 2J & 47 & $46\pm 7$ & 0 & 1.2 & 0 & 3.0 & 0 & 2.9 & 0 & 1.6 & 0 & 2.3 & 0 & 1.8 & 0\\
GG 6J bulk & 32 & $24\pm 5$ & 1.3 & 6.0 & 0.1 & 5.7 & 0.1 & 4.3 & 0.3 & 4.6 & 0.3 & 6.6 & 0.1 & 6.9 & 0 \\
GG 6J high-mass & 3 & $3.8\pm 1.2$ & 0 & 1.0 & 0.4 & 3.2 & 2.2 & 1.4 & 0.7 & 1.1 & 0.5 & 2.4 & 1.4 & 1.9 & 1.0 \\
GG 4J low-x & 4 & $6.0\pm 1.6$ & 0.4 & 0.9 & 0.8 & 4.2 & 3.7 & 1.9 & 1.5 & 1.5 & 1.2 & 1.9 & 1.5 & 1.8 & 1.5\\
GG 4J low-x b-veto & 2 & $3.3\pm 1.2$ & 0.2 & 0.1 & 0.3 & 0 & 0.2 & 0.1 & 0.3 & 0.1 & 0.3& 0.4 & 0.5 & 0 & 0.2\\
GG 4J high-x & 2 & $3.4\pm 0.9$ & 0.4 & 0.5 & 0.7 & 0.4 & 0.7 & 0.4 & 0.7 & 0.3 & 0.6 & 1.0 & 1.2 & 0.9 & 1.1\\
SS 4J x=1/2 & 6 & $5.4\pm 1.7$ & 0.1 & 7.1 & 3.2 & 9.5 & 5.6 & 6.8 & 3.0 & 5.8 & 2.1 & 7.9 & 4.0 & 9.2 & 5.3\\
SS 5J x=1/2 & 8 & $13.2\pm 2.5$ & 1.4 & 0.6 & 1.8 & 0.4 & 1.6 & 0.5 & 1.7 & 0.3 & 1.6 & 0.9 & 1.9 & 0.7 & 1.8 \\
SS 4J low-x & 8 & $11.1\pm 2.7$ & 0.4 & 1.3 & 0.9 & 3.2 & 1.9 & 2.8 & 1.7 & 1.9 & 1.2 & 1.5 & 1.0 & 4.5 & 2.8 \\
SS 5J high-x & 7 & $4.6\pm 1.4$ & 0.9 & 2.4 & 0 & 3.4 & 0.1 & 1.1 & 0.3 & 2.5 & 0 & 2.1 & 0 & 2.1 & 0\\
 \hline
      \hline
 \multicolumn{3}{|c||}{Total $\Delta\chi^2$}& 1.4 &  \multicolumn{2}{|c|}{3.2} & \multicolumn{2}{|c|}{5.6} & \multicolumn{2}{|c|}{3.0} & \multicolumn{2}{|c|}{2.1} & \multicolumn{2}{|c|}{4.0} &\multicolumn{2}{|c|}{5.3} \\
\hline
 \end{tabular}
 \caption{Breakdown of the $\chi^2$ contributions due to different signal bins of the ATLAS 1 lepton + jets search\cite{ATLAS-CONF-2016-054} for the background-only hypothesis (SM) and the benchmark points. The total $\Delta\chi^2$ is given by the maximum of the individual contributions. The bins are labeled according to their ATLAS name. }
\label{tab:chisq_ATLAS_1l}
\end{table}
%%%%%%%%%%%%%%%%%%%%%%%%%%%%%%%%%%%%%%%%%%%%%%%%%%%%%%%%%

%%%%%%%%%%%%%%%%%%%%%%%%%%%%%%%%%%%%%%%%%%%%%%%%%%%%%%%
\begin{table}[H]\footnotesize
 \centering
 \begin{tabular}{|c|c|c||c|cc|cc|cc|cc|cc|cc|} 
 \hline
 Bin & Obs. & Bkg. &  SM & \multicolumn{2}{c|}{BP1} & \multicolumn{2}{c|}{BP2} & \multicolumn{2}{c|}{BP3} & \multicolumn{2}{c|}{BP4} & \multicolumn{2}{c|}{BP5} & \multicolumn{2}{c|}{BP6} \\
& & & $\Delta\chi^2_i$ & $s_i$ & $\Delta\chi^2_i$ & $s_i$ & $\Delta\chi^2_i$ & $s_i$ & $\Delta\chi^2_i$ & $s_i$ & $\Delta\chi^2_i$ & $s_i$ & $\Delta\chi^2_i$ & $s_i$ & $\Delta\chi^2_i$\\
 \hline
 \hline
SRA250 & 23 & $29 \pm 5$ & 0.6 & 0.8 & 0.8 & 1.8 & 1.1 & 1.0 & 0.8 & 3.0 & 1.4 & 0.4 & 0.7 & 1.2 & 0.9 \\
SRA350 & 6 & $7.0 \pm 1.2$ & 0.1 & 0.6 & 0.2 & 1.1 & 0.4 & 0.6 & 0.2 & 2.0 & 0.9 & 0.3 & 0.2 & 0.8 & 0.3\\
SRA450 & 1 & $1.8 \pm 0.4$ & 0.3 & 0.4 & 0.7 & 0.4 & 0.7 & 0.2 & 0.5 & 1.2 & 1.6 & 0.1 & 0.4 & 0.4 & 0.7\\
SRB & 6 & $12.0 \pm 2.5$  & 2.2 & 0.1 & 2.3 & 0.2 & 2.4 & 0.2 & 2.4 & 0.4 & 2.5 & 0.2 & 2.4 & 0.4 & 2.5\\
 \hline
       \hline
 \multicolumn{3}{|c||}{Total $\Delta\chi^2$}& 2.2 &  \multicolumn{2}{|c|}{2.3} & \multicolumn{2}{|c|}{2.4} & \multicolumn{2}{|c|}{2.4} & \multicolumn{2}{|c|}{2.5} & \multicolumn{2}{|c|}{2.4} &\multicolumn{2}{|c|}{2.5} \\
\hline
 \end{tabular}
 \caption{Breakdown of the $\chi^2$ contributions due to different signal bins of the ATLAS 2 b-tagged jets search\cite{Aaboud:2016nwl} for the background-only hypothesis (SM) and the benchmark points. The total $\Delta\chi^2$ is given by the maximum of the individual contributions. The bins are labeled according to their ATLAS name. }
\label{tab:chisq_ATLAS_2bj}
\end{table}
%%%%%%%%%%%%%%%%%%%%%%%%%%%%%%%%%%%%%%%%%%%%%%%%%%%%%%%%%

%%%%%%%%%%%%%%%%%%%%%%%%%%%%%%%%%%%%%%%%%%%%%%%%%%%%%%%
\bibliographystyle{utphysmcite}	% (uses file "plain.bst")
\bibliography{myref}

\ifx\mcitethebibliography\mciteundefinedmacro
\PackageError{unsrtM.bst}{mciteplus.sty has not been loaded}
{This bibstyle requires the use of the mciteplus package.}\fi
\begin{mcitethebibliography}{10}

\bibitem{ichep16}
{\url{http://indico.cern.ch/event/432527/timetable/#20160804}}\relax
\mciteBstWouldAddEndPunctfalse
\mciteSetBstMidEndSepPunct{\mcitedefaultmidpunct}
{}{\mcitedefaultseppunct}\relax
\EndOfBibitem
\bibitem{ATLAS-CONF-2016-050}
{\bfseries ATLAS} Collaboration, ``{Search for top squarks in final states with
  one isolated lepton, jets, and missing transverse momentum in $\sqrt{s}$ = 13
  TeV pp collisions with the ATLAS detector},'' Tech. Rep. ATLAS-CONF-2016-050,
  CERN, Geneva, Aug, 2016.
\newblock \url{https://cds.cern.ch/record/2206132}\relax
\mciteBstWouldAddEndPunctfalse
\mciteSetBstMidEndSepPunct{\mcitedefaultmidpunct}
{}{\mcitedefaultseppunct}\relax
\EndOfBibitem
\bibitem{Berger:2008cq}
C.~F. Berger, J.~S. Gainer, J.~L. Hewett, and T.~G. Rizzo, ``{Supersymmetry
  Without Prejudice},''
  \href{http://dx.doi.org/10.1088/1126-6708/2009/02/023}{{\em JHEP} {\bfseries
  02} (2009) 023},
\href{http://arxiv.org/abs/0812.0980}{{\ttfamily arXiv:0812.0980 [hep-ph]}}.
%%CITATION = ARXIV:0812.0980;%%\relax
\mciteBstWouldAddEndPunctfalse
\mciteSetBstMidEndSepPunct{\mcitedefaultmidpunct}
{}{\mcitedefaultseppunct}\relax
\EndOfBibitem
\bibitem{Aad:2015baa}
{\bfseries ATLAS} Collaboration, G.~Aad {\em et~al.}, ``{Summary of the ATLAS
  experiment’s sensitivity to supersymmetry after LHC Run 1 — interpreted
  in the phenomenological MSSM},''
  \href{http://dx.doi.org/10.1007/JHEP10(2015)134}{{\em JHEP} {\bfseries 10}
  (2015) 134},
\href{http://arxiv.org/abs/1508.06608}{{\ttfamily arXiv:1508.06608 [hep-ex]}}.
%%CITATION = ARXIV:1508.06608;%%\relax
\mciteBstWouldAddEndPunctfalse
\mciteSetBstMidEndSepPunct{\mcitedefaultmidpunct}
{}{\mcitedefaultseppunct}\relax
\EndOfBibitem
\bibitem{Kowalska:2016ent}
K.~Kowalska, ``{Phenomenological MSSM in light of new 13 TeV LHC data},''
  \href{http://dx.doi.org/10.1140/epjc/s10052-016-4536-4}{{\em Eur. Phys. J.}
  {\bfseries C76} no.~12, (2016) 684},
\href{http://arxiv.org/abs/1608.02489}{{\ttfamily arXiv:1608.02489 [hep-ph]}}.
%%CITATION = ARXIV:1608.02489;%%\relax
\mciteBstWouldAddEndPunctfalse
\mciteSetBstMidEndSepPunct{\mcitedefaultmidpunct}
{}{\mcitedefaultseppunct}\relax
\EndOfBibitem
\bibitem{CMS-PAS-SUS-16-028}
{\bfseries CMS} Collaboration, ``{Search for direct top squark pair production
  in the single lepton final state at $\sqrt{s}=13~\mathrm{TeV}$},'' Tech. Rep.
  CMS-PAS-SUS-16-028, CERN, Geneva, 2016.
\newblock \url{http://cds.cern.ch/record/2205271}\relax
\mciteBstWouldAddEndPunctfalse
\mciteSetBstMidEndSepPunct{\mcitedefaultmidpunct}
{}{\mcitedefaultseppunct}\relax
\EndOfBibitem
\bibitem{Han:2016hgr}
C.~Han, M.~M. Nojiri, M.~Takeuchi, and T.~T. Yanagida, ``{Surviving scenario of
  stop decays for ATLAS $\ell+jets+E^{miss}_T$ search},''
\href{http://arxiv.org/abs/1609.09303}{{\ttfamily arXiv:1609.09303 [hep-ph]}}.
%%CITATION = ARXIV:1609.09303;%%\relax
\mciteBstWouldAddEndPunctfalse
\mciteSetBstMidEndSepPunct{\mcitedefaultmidpunct}
{}{\mcitedefaultseppunct}\relax
\EndOfBibitem
\bibitem{CMS-PAS-SUS-16-029}
{\bfseries CMS} Collaboration, ``{Search for direct top squark pair production
  in the fully hadronic final state in proton-proton collisions at sqrt(s) = 13
  TeV corresponding to an integrated luminosity of 12.9/fb},'' Tech. Rep.
  CMS-PAS-SUS-16-029, CERN, Geneva, 2016.
\newblock \url{https://cds.cern.ch/record/2205176}\relax
\mciteBstWouldAddEndPunctfalse
\mciteSetBstMidEndSepPunct{\mcitedefaultmidpunct}
{}{\mcitedefaultseppunct}\relax
\EndOfBibitem
\bibitem{CMS-PAS-SUS-16-030}
{\bfseries CMS} Collaboration, ``{Search for supersymmetry in the all-hadronic
  final state using top quark tagging in pp collisions at sqrt(s) = 13 TeV},''
  Tech. Rep. CMS-PAS-SUS-16-030, CERN, Geneva, 2016.
\newblock \url{https://cds.cern.ch/record/2204930}\relax
\mciteBstWouldAddEndPunctfalse
\mciteSetBstMidEndSepPunct{\mcitedefaultmidpunct}
{}{\mcitedefaultseppunct}\relax
\EndOfBibitem
\bibitem{CahillRowley:2012cb}
M.~W. Cahill-Rowley, J.~L. Hewett, S.~Hoeche, A.~Ismail, and T.~G. Rizzo,
  ``{The New Look pMSSM with Neutralino and Gravitino LSPs},''
  \href{http://dx.doi.org/10.1140/epjc/s10052-012-2156-1}{{\em Eur. Phys. J.}
  {\bfseries C72} (2012) 2156},
\href{http://arxiv.org/abs/1206.4321}{{\ttfamily arXiv:1206.4321 [hep-ph]}}.
%%CITATION = ARXIV:1206.4321;%%\relax
\mciteBstWouldAddEndPunctfalse
\mciteSetBstMidEndSepPunct{\mcitedefaultmidpunct}
{}{\mcitedefaultseppunct}\relax
\EndOfBibitem
\bibitem{CahillRowley:2012kx}
M.~W. Cahill-Rowley, J.~L. Hewett, A.~Ismail, and T.~G. Rizzo, ``{More energy,
  more searches, but the phenomenological MSSM lives on},''
  \href{http://dx.doi.org/10.1103/PhysRevD.88.035002}{{\em Phys. Rev.}
  {\bfseries D88} no.~3, (2013) 035002},
\href{http://arxiv.org/abs/1211.1981}{{\ttfamily arXiv:1211.1981 [hep-ph]}}.
%%CITATION = ARXIV:1211.1981;%%\relax
\mciteBstWouldAddEndPunctfalse
\mciteSetBstMidEndSepPunct{\mcitedefaultmidpunct}
{}{\mcitedefaultseppunct}\relax
\EndOfBibitem
\bibitem{Cahill-Rowley:2014twa}
M.~Cahill-Rowley, J.~L. Hewett, A.~Ismail, and T.~G. Rizzo, ``{Lessons and
  prospects from the pMSSM after LHC Run I},''
  \href{http://dx.doi.org/10.1103/PhysRevD.91.055002}{{\em Phys. Rev.}
  {\bfseries D91} no.~5, (2015) 055002},
\href{http://arxiv.org/abs/1407.4130}{{\ttfamily arXiv:1407.4130 [hep-ph]}}.
%%CITATION = ARXIV:1407.4130;%%\relax
\mciteBstWouldAddEndPunctfalse
\mciteSetBstMidEndSepPunct{\mcitedefaultmidpunct}
{}{\mcitedefaultseppunct}\relax
\EndOfBibitem
\bibitem{Allanach:2001kg}
B.~Allanach, ``{SOFTSUSY: a program for calculating supersymmetric spectra},''
  \href{http://dx.doi.org/10.1016/S0010-4655(01)00460-X}{{\em
  Comput.Phys.Commun.} {\bfseries 143} (2002) 305--331},
\href{http://arxiv.org/abs/hep-ph/0104145}{{\ttfamily arXiv:hep-ph/0104145
  [hep-ph]}}.
%%CITATION = HEP-PH/0104145;%%\relax
\mciteBstWouldAddEndPunctfalse
\mciteSetBstMidEndSepPunct{\mcitedefaultmidpunct}
{}{\mcitedefaultseppunct}\relax
\EndOfBibitem
\bibitem{Djouadi:2006bz}
A.~Djouadi, M.~M. Muhlleitner, and M.~Spira, ``{Decays of supersymmetric
  particles: The Program SUSY-HIT (SUspect-SdecaY-Hdecay-InTerface)},'' {\em
  Acta Phys. Polon.} {\bfseries B38} (2007) 635--644,
\href{http://arxiv.org/abs/hep-ph/0609292}{{\ttfamily arXiv:hep-ph/0609292
  [hep-ph]}}.
%%CITATION = HEP-PH/0609292;%%\relax
\mciteBstWouldAddEndPunctfalse
\mciteSetBstMidEndSepPunct{\mcitedefaultmidpunct}
{}{\mcitedefaultseppunct}\relax
\EndOfBibitem
\bibitem{Belanger:2006is}
G.~Belanger, F.~Boudjema, A.~Pukhov, and A.~Semenov, ``{MicrOMEGAs 2.0: A
  Program to calculate the relic density of dark matter in a generic model},''
  \href{http://dx.doi.org/10.1016/j.cpc.2006.11.008}{{\em Comput. Phys.
  Commun.} {\bfseries 176} (2007) 367--382},
\href{http://arxiv.org/abs/hep-ph/0607059}{{\ttfamily arXiv:hep-ph/0607059
  [hep-ph]}}.
%%CITATION = HEP-PH/0607059;%%\relax
\mciteBstWouldAddEndPunctfalse
\mciteSetBstMidEndSepPunct{\mcitedefaultmidpunct}
{}{\mcitedefaultseppunct}\relax
\EndOfBibitem
\bibitem{Belanger:2010pz}
G.~Belanger, F.~Boudjema, A.~Pukhov, and A.~Semenov, ``{micrOMEGAs: A Tool for
  dark matter studies},''
  \href{http://dx.doi.org/10.1393/ncc/i2010-10591-3}{{\em Nuovo Cim.}
  {\bfseries C033N2} (2010) 111--116},
\href{http://arxiv.org/abs/1005.4133}{{\ttfamily arXiv:1005.4133 [hep-ph]}}.
%%CITATION = ARXIV:1005.4133;%%\relax
\mciteBstWouldAddEndPunctfalse
\mciteSetBstMidEndSepPunct{\mcitedefaultmidpunct}
{}{\mcitedefaultseppunct}\relax
\EndOfBibitem
\bibitem{Mahmoudi:2008tp}
F.~Mahmoudi, ``{SuperIso v2.3: A Program for calculating flavor physics
  observables in Supersymmetry},''
  \href{http://dx.doi.org/10.1016/j.cpc.2009.02.017}{{\em Comput. Phys.
  Commun.} {\bfseries 180} (2009) 1579--1613},
\href{http://arxiv.org/abs/0808.3144}{{\ttfamily arXiv:0808.3144 [hep-ph]}}.
%%CITATION = ARXIV:0808.3144;%%\relax
\mciteBstWouldAddEndPunctfalse
\mciteSetBstMidEndSepPunct{\mcitedefaultmidpunct}
{}{\mcitedefaultseppunct}\relax
\EndOfBibitem
\bibitem{Heinemeyer:1998yj}
S.~Heinemeyer, W.~Hollik, and G.~Weiglein, ``{FeynHiggs: A Program for the
  calculation of the masses of the neutral CP even Higgs bosons in the MSSM},''
  \href{http://dx.doi.org/10.1016/S0010-4655(99)00364-1}{{\em Comput. Phys.
  Commun.} {\bfseries 124} (2000) 76--89},
\href{http://arxiv.org/abs/hep-ph/9812320}{{\ttfamily arXiv:hep-ph/9812320
  [hep-ph]}}.
%%CITATION = HEP-PH/9812320;%%\relax
\mciteBstWouldAddEndPunctfalse
\mciteSetBstMidEndSepPunct{\mcitedefaultmidpunct}
{}{\mcitedefaultseppunct}\relax
\EndOfBibitem
\bibitem{Hahn:2013ria}
T.~Hahn, S.~Heinemeyer, W.~Hollik, H.~Rzehak, and G.~Weiglein,
  ``{High-Precision Predictions for the Light CP -Even Higgs Boson Mass of the
  Minimal Supersymmetric Standard Model},''
  \href{http://dx.doi.org/10.1103/PhysRevLett.112.141801}{{\em Phys. Rev.
  Lett.} {\bfseries 112} no.~14, (2014) 141801},
\href{http://arxiv.org/abs/1312.4937}{{\ttfamily arXiv:1312.4937 [hep-ph]}}.
%%CITATION = ARXIV:1312.4937;%%\relax
\mciteBstWouldAddEndPunctfalse
\mciteSetBstMidEndSepPunct{\mcitedefaultmidpunct}
{}{\mcitedefaultseppunct}\relax
\EndOfBibitem
\bibitem{Han:2016xet}
C.~Han, J.~Ren, L.~Wu, J.~M. Yang, and M.~Zhang, ``{Top-squark in natural SUSY
  under current LHC run-2 data},''
\href{http://arxiv.org/abs/1609.02361}{{\ttfamily arXiv:1609.02361 [hep-ph]}}.
%%CITATION = ARXIV:1609.02361;%%\relax
\mciteBstWouldAddEndPunctfalse
\mciteSetBstMidEndSepPunct{\mcitedefaultmidpunct}
{}{\mcitedefaultseppunct}\relax
\EndOfBibitem
\bibitem{Buckley:2016kvr}
M.~R. Buckley, D.~Feld, S.~Macaluso, A.~Monteux, and D.~Shih, ``{Cornering
  Natural SUSY at LHC Run II and Beyond},''
\href{http://arxiv.org/abs/1610.08059}{{\ttfamily arXiv:1610.08059 [hep-ph]}}.
%%CITATION = ARXIV:1610.08059;%%\relax
\mciteBstWouldAddEndPunctfalse
\mciteSetBstMidEndSepPunct{\mcitedefaultmidpunct}
{}{\mcitedefaultseppunct}\relax
\EndOfBibitem
\bibitem{Aaboud:2016zdn}
{\bfseries ATLAS} Collaboration, M.~Aaboud {\em et~al.}, ``{Search for squarks
  and gluinos in final states with jets and missing transverse momentum at
  $\sqrt{s} =$ 13 TeV with the ATLAS detector},''
  \href{http://dx.doi.org/10.1140/epjc/s10052-016-4184-8}{{\em Eur. Phys. J.}
  {\bfseries C76} no.~7, (2016) 392},
\href{http://arxiv.org/abs/1605.03814}{{\ttfamily arXiv:1605.03814 [hep-ex]}}.
%%CITATION = ARXIV:1605.03814;%%\relax
\mciteBstWouldAddEndPunctfalse
\mciteSetBstMidEndSepPunct{\mcitedefaultmidpunct}
{}{\mcitedefaultseppunct}\relax
\EndOfBibitem
\bibitem{ATLAS-CONF-2016-078}
{\bfseries ATLAS} Collaboration, ``{Further searches for squarks and gluinos in
  final states with jets and missing transverse momentum at $\sqrt{s}$ =13 TeV
  with the ATLAS detector},'' Tech. Rep. ATLAS-CONF-2016-078, CERN, Geneva,
  Aug, 2016.
\newblock \url{https://cds.cern.ch/record/2206252}\relax
\mciteBstWouldAddEndPunctfalse
\mciteSetBstMidEndSepPunct{\mcitedefaultmidpunct}
{}{\mcitedefaultseppunct}\relax
\EndOfBibitem
\bibitem{Aad:2016qqk}
{\bfseries ATLAS} Collaboration, G.~Aad {\em et~al.}, ``{Search for gluinos in
  events with an isolated lepton, jets and missing transverse momentum at
  $\sqrt{s}$ = 13 TeV with the ATLAS detector},''
  \href{http://dx.doi.org/10.1140/epjc/s10052-016-4397-x}{{\em Eur. Phys. J.}
  {\bfseries C76} no.~10, (2016) 565},
\href{http://arxiv.org/abs/1605.04285}{{\ttfamily arXiv:1605.04285 [hep-ex]}}.
%%CITATION = ARXIV:1605.04285;%%\relax
\mciteBstWouldAddEndPunctfalse
\mciteSetBstMidEndSepPunct{\mcitedefaultmidpunct}
{}{\mcitedefaultseppunct}\relax
\EndOfBibitem
\bibitem{ATLAS-CONF-2016-054}
{\bfseries ATLAS} Collaboration, ``{Search for squarks and gluinos in events
  with an isolated lepton, jets and missing transverse momentum at $\sqrt{s}$ =
  13 TeV with the ATLAS detector},'' Tech. Rep. ATLAS-CONF-2016-054, CERN,
  Geneva, Aug, 2016.
\newblock \url{https://cds.cern.ch/record/2206136}\relax
\mciteBstWouldAddEndPunctfalse
\mciteSetBstMidEndSepPunct{\mcitedefaultmidpunct}
{}{\mcitedefaultseppunct}\relax
\EndOfBibitem
\bibitem{Aad:2016eki}
{\bfseries ATLAS} Collaboration, G.~Aad {\em et~al.}, ``{Search for pair
  production of gluinos decaying via stop and sbottom in events with $b$-jets
  and large missing transverse momentum in $pp$ collisions at $\sqrt{s} = 13$
  TeV with the ATLAS detector},''
  \href{http://dx.doi.org/10.1103/PhysRevD.94.032003}{{\em Phys. Rev.}
  {\bfseries D94} no.~3, (2016) 032003},
\href{http://arxiv.org/abs/1605.09318}{{\ttfamily arXiv:1605.09318 [hep-ex]}}.
%%CITATION = ARXIV:1605.09318;%%\relax
\mciteBstWouldAddEndPunctfalse
\mciteSetBstMidEndSepPunct{\mcitedefaultmidpunct}
{}{\mcitedefaultseppunct}\relax
\EndOfBibitem
\bibitem{ATLAS-CONF-2016-052}
{\bfseries ATLAS} Collaboration, ``{Search for pair production of gluinos
  decaying via top or bottom squarks in events with $b$-jets and large missing
  transverse momentum in $pp$ collisions at $\sqrt{s}=13$ TeV with the ATLAS
  detector},'' Tech. Rep. ATLAS-CONF-2016-052, CERN, Geneva, Aug, 2016.
\newblock \url{https://cds.cern.ch/record/2206134}\relax
\mciteBstWouldAddEndPunctfalse
\mciteSetBstMidEndSepPunct{\mcitedefaultmidpunct}
{}{\mcitedefaultseppunct}\relax
\EndOfBibitem
\bibitem{ATLAS-CONF-2016-077}
{\bfseries ATLAS} Collaboration, ``{Search for the Supersymmetric Partner of
  the Top Quark in the Jets+Emiss Final State at sqrt(s) = 13 TeV},'' Tech.
  Rep. ATLAS-CONF-2016-077, CERN, Geneva, Aug, 2016.
\newblock \url{https://cds.cern.ch/record/2206250}\relax
\mciteBstWouldAddEndPunctfalse
\mciteSetBstMidEndSepPunct{\mcitedefaultmidpunct}
{}{\mcitedefaultseppunct}\relax
\EndOfBibitem
\bibitem{Aaboud:2016lwz}
{\bfseries ATLAS} Collaboration, M.~Aaboud {\em et~al.}, ``{Search for top
  squarks in final states with one isolated lepton, jets, and missing
  transverse momentum in $\sqrt{s}=13$ TeV $pp$ collisions with the ATLAS
  detector},'' \href{http://dx.doi.org/10.1103/PhysRevD.94.052009}{{\em Phys.
  Rev.} {\bfseries D94} no.~5, (2016) 052009},
\href{http://arxiv.org/abs/1606.03903}{{\ttfamily arXiv:1606.03903 [hep-ex]}}.
%%CITATION = ARXIV:1606.03903;%%\relax
\mciteBstWouldAddEndPunctfalse
\mciteSetBstMidEndSepPunct{\mcitedefaultmidpunct}
{}{\mcitedefaultseppunct}\relax
\EndOfBibitem
\bibitem{ATLAS-CONF-2016-009}
``{Search for direct top squark pair production in final states with two
  leptons in $\sqrt{s} = 13$ TeV pp collisions using 3.2 fb$^{-1}$ of ATLAS
  data.},'' Tech. Rep. ATLAS-CONF-2016-009, CERN, Geneva, Mar, 2016.
\newblock \url{https://cds.cern.ch/record/2139643}\relax
\mciteBstWouldAddEndPunctfalse
\mciteSetBstMidEndSepPunct{\mcitedefaultmidpunct}
{}{\mcitedefaultseppunct}\relax
\EndOfBibitem
\bibitem{Aaboud:2016nwl}
{\bfseries ATLAS} Collaboration, M.~Aaboud {\em et~al.}, ``{Search for bottom
  squark pair production in proton–proton collisions at $\sqrt{s}=13$ TeV
  with the ATLAS detector},''
  \href{http://dx.doi.org/10.1140/epjc/s10052-016-4382-4}{{\em Eur. Phys. J.}
  {\bfseries C76} no.~10, (2016) 547},
\href{http://arxiv.org/abs/1606.08772}{{\ttfamily arXiv:1606.08772 [hep-ex]}}.
%%CITATION = ARXIV:1606.08772;%%\relax
\mciteBstWouldAddEndPunctfalse
\mciteSetBstMidEndSepPunct{\mcitedefaultmidpunct}
{}{\mcitedefaultseppunct}\relax
\EndOfBibitem
\bibitem{Aaboud:2016tnv}
{\bfseries ATLAS} Collaboration, M.~Aaboud {\em et~al.}, ``{Search for new
  phenomena in final states with an energetic jet and large missing transverse
  momentum in $pp$ collisions at $\sqrt{s}=13$  TeV using the ATLAS
  detector},'' \href{http://dx.doi.org/10.1103/PhysRevD.94.032005}{{\em Phys.
  Rev.} {\bfseries D94} no.~3, (2016) 032005},
\href{http://arxiv.org/abs/1604.07773}{{\ttfamily arXiv:1604.07773 [hep-ex]}}.
%%CITATION = ARXIV:1604.07773;%%\relax
\mciteBstWouldAddEndPunctfalse
\mciteSetBstMidEndSepPunct{\mcitedefaultmidpunct}
{}{\mcitedefaultseppunct}\relax
\EndOfBibitem
\bibitem{CMS-PAS-SUS-16-014}
{\bfseries CMS} Collaboration, ``{Search for supersymmetry in events with jets
  and missing transverse momentum in proton-proton collisions at 13 TeV},''
  Tech. Rep. CMS-PAS-SUS-16-014, CERN, Geneva, 2016.
\newblock \url{https://cds.cern.ch/record/2205158}\relax
\mciteBstWouldAddEndPunctfalse
\mciteSetBstMidEndSepPunct{\mcitedefaultmidpunct}
{}{\mcitedefaultseppunct}\relax
\EndOfBibitem
\bibitem{CMS-PAS-SUS-16-015}
{\bfseries CMS} Collaboration, ``{Search for new physics in the all-hadronic
  final state with the MT2 variable},'' Tech. Rep. CMS-PAS-SUS-16-015, CERN,
  Geneva, 2016.
\newblock \url{https://cds.cern.ch/record/2205162}\relax
\mciteBstWouldAddEndPunctfalse
\mciteSetBstMidEndSepPunct{\mcitedefaultmidpunct}
{}{\mcitedefaultseppunct}\relax
\EndOfBibitem
\bibitem{CMS-PAS-SUS-16-016}
{\bfseries CMS} Collaboration, ``{An inclusive search for new phenomena in
  final states with one or more jets and missing transverse momentum at 13 TeV
  with the AlphaT variable},'' Tech. Rep. CMS-PAS-SUS-16-016, CERN, Geneva,
  2016.
\newblock \url{https://cds.cern.ch/record/2205163}\relax
\mciteBstWouldAddEndPunctfalse
\mciteSetBstMidEndSepPunct{\mcitedefaultmidpunct}
{}{\mcitedefaultseppunct}\relax
\EndOfBibitem
\bibitem{Sjostrand:2007gs}
T.~Sjostrand, S.~Mrenna, and P.~Z. Skands, ``{A Brief Introduction to PYTHIA
  8.1},'' \href{http://dx.doi.org/10.1016/j.cpc.2008.01.036}{{\em
  Comput.Phys.Commun.} {\bfseries 178} (2008) 852--867},
\href{http://arxiv.org/abs/0710.3820}{{\ttfamily arXiv:0710.3820 [hep-ph]}}.
%%CITATION = ARXIV:0710.3820;%%\relax
\mciteBstWouldAddEndPunctfalse
\mciteSetBstMidEndSepPunct{\mcitedefaultmidpunct}
{}{\mcitedefaultseppunct}\relax
\EndOfBibitem
\bibitem{deFavereau:2013fsa}
{\bfseries DELPHES 3} Collaboration, J.~de~Favereau {\em et~al.}, ``{DELPHES 3,
  A modular framework for fast simulation of a generic collider experiment},''
  \href{http://dx.doi.org/10.1007/JHEP02(2014)057}{{\em JHEP} {\bfseries 1402}
  (2014) 057},
\href{http://arxiv.org/abs/1307.6346}{{\ttfamily arXiv:1307.6346 [hep-ex]}}.
%%CITATION = ARXIV:1307.6346;%%\relax
\mciteBstWouldAddEndPunctfalse
\mciteSetBstMidEndSepPunct{\mcitedefaultmidpunct}
{}{\mcitedefaultseppunct}\relax
\EndOfBibitem
\bibitem{Chacko:2001km}
Z.~Chacko and E.~Ponton, ``{Yukawa deflected gauge mediation},''
  \href{http://dx.doi.org/10.1103/PhysRevD.66.095004}{{\em Phys. Rev.}
  {\bfseries D66} (2002) 095004},
\href{http://arxiv.org/abs/hep-ph/0112190}{{\ttfamily arXiv:hep-ph/0112190
  [hep-ph]}}.
%%CITATION = HEP-PH/0112190;%%\relax
\mciteBstWouldAddEndPunctfalse
\mciteSetBstMidEndSepPunct{\mcitedefaultmidpunct}
{}{\mcitedefaultseppunct}\relax
\EndOfBibitem
\bibitem{Bai:2012gs}
Y.~Bai, H.-C. Cheng, J.~Gallicchio, and J.~Gu, ``{Stop the Top Background of
  the Stop Search},'' \href{http://dx.doi.org/10.1007/JHEP07(2012)110}{{\em
  JHEP} {\bfseries 07} (2012) 110},
\href{http://arxiv.org/abs/1203.4813}{{\ttfamily arXiv:1203.4813 [hep-ph]}}.
%%CITATION = ARXIV:1203.4813;%%\relax
\mciteBstWouldAddEndPunctfalse
\mciteSetBstMidEndSepPunct{\mcitedefaultmidpunct}
{}{\mcitedefaultseppunct}\relax
\EndOfBibitem
\end{mcitethebibliography}

%%%%%%%%%%%%%%%%%%%%%%%%%%%%%%%%%%%%%%%%%%%%%%%%%%%%

\end{document}